\documentclass[aps,twocolumn,groupedaddress,showpacs,amsmath,amssymb,pra]{revtex4-1}

\usepackage{graphicx}
\usepackage{array}
\usepackage{mdwmath}
\usepackage{color}
\usepackage{siunitx}
\usepackage{amsmath}
\usepackage{amssymb}
\usepackage{pdfpages}
\allowdisplaybreaks
\def\ptl{IEEE\ Photon.\ Technol.\ Lett.\ }

\def\opex{Opt.\ Express\ }

\def\el{Electron.\ Lett.\ }
\def\aop{Adv.\ Opt.\ Photon.\ }

\def\tmtt{IEEE\ Trans.\ Microw.\ Theory\ Tech.\ }

\def\jstqe{IEEE\ J.\ Sel.\ Top.\ Quantum\ Electron.\ }
\def\sci{Science\ }
\def\nphot{Nat.\ Photonics\ }
\def\nm{Nat.\ Mater.\ }

\begin{document}

\title{Theory of Pulsed Four-Wave-Mixing in One-dimensional Silicon Photonic Crystal Slab Waveguides}

\author{Spyros Lavdas}
\author{Nicolae C. Panoiu}
\affiliation{Department of Electronic and Electrical Engineering, University College London,
Torrington Place, London WC1E 7JE, United Kingdom}

\date{\today}

\begin{abstract}
We present a comprehensive theoretical analysis and computational study of four-wave mixing (FWM)
of optical pulses co-propagating in one-dimensional silicon photonic crystal waveguides
(Si-PhCWGs). Our theoretical analysis describes a very general set-up of the interacting optical
pulses, namely we consider nondegenerate FWM in a configuration in which at each frequency there
exists a superposition of guiding modes. We incorporate in our theoretical model all relevant
linear optical effects, including waveguide loss, free-carrier (FC) dispersion and FC absorption,
nonlinear optical effects such as self- and cross-phase modulation (SPM, XPM), two-photon
absorption (TPA), and cross-absorption modulation (XAM), as well as the coupled dynamics of FCs
and optical field. In particular, our theoretical analysis based on the coupled-mode theory
provides rigorously derived formulae for linear dispersion coefficients of the guiding modes,
linear coupling coefficients between these modes, as well as the nonlinear waveguide coefficients
describing SPM, XPM, TPA, XAM, and FWM. In addition, our theoretical analysis and numerical
simulations reveal key differences between the characteristics of FWM in the slow- and fast-light
regimes, which could potentially have important implications to the design of ultra-compact active
photonic devices.
\end{abstract}


\maketitle
\section{Introduction}\label{sIntr}
One of the most promising applications of photonics is the development of ultra-compact optical
interconnects for chip-to-chip and even intra-chip communications. The driving forces behind
research in this area are the perceived limitations at high frequency of currently used copper
interconnects \cite{hmh01pieee}, combined with a rapidly increasing demand to move huge amounts of
data within increasingly more confined yet increasingly intricate communication architectures. An
approach showing great potential towards developing optical interconnects at chip scale is based
on high-index contrast optical waveguides, such as silicon photonic waveguides (Si-PhWGs)
implemented on the silicon-on-insulator material platform \cite{lll00apl,apc02ptl}. Among key
advantages provided by this platform are the increased potential for device integration
facilitated by the enhanced confinement of the optical field achievable in high-index contrast
photonic structures, as well as the particularly large optical nonlinearity of silicon, which
makes it an ideal material for active photonic devices. Many of the basic device functionalities
required in networks-on-chip have in fact already been demonstrated using Si-PhWGs, including
parametric amplification \cite{cdr03oe,edo04oe,rjl05n,fts06n,lov10np}, optical modulation
\cite{cir95ptl,ljl04n,xsp05n}, pulse compression \cite{cph06ptl,m08ptl}, supercontinuum generation
\cite{bkr04oe,hcl07oe}, pulse self-steepening \cite{plo09ol}, modulational instability
\cite{pco06ol}, and four-wave mixing (FWM) \cite{fys05oe,edo05oe,fts07oe,zpm10np,dog12oe}; for a
review of optical properties of Si-PhWGs see \cite{opd09aop}. However, since the parameter space
of Si-PhWGs is rather limited, there is little room to engineer their optical properties.

A promising solution to this problem has its roots in the advent of photonic crystals (PhCs) in
the late 80's \cite{y87,j87}. Thus, by patterning an optical medium in a periodic manner, with the
spatial periods of the pattern being comparable to the operating optical wavelength, the optical
properties of the resulting medium can be modified and engineered to a remarkable extent.
Following this approach, a series of photonic devices have been demonstrated using PhCs, including
optical waveguides and bends \cite{mck96prl,lch98s,bfy99el,mck99prb,cn00prb}, optical
micro-cavities \cite{fvf97nat,pls99s,aas03nat,rsl04nat,ysh04nat}, and optical filters
\cite{fvj99prb,cmi01apl}. One of the most effective approaches to affect the optical properties of
PhCs is to modify the group-velocity (GV), $v_{g}$, of the propagating modes. Unlike the case of
waves propagating in regular optical media, whose GV can hardly be altered, by varying the
geometrical parameters of PhCs one can tune the corresponding GV over many orders of magnitude.
Perhaps the most noteworthy implication of the existence of optical modes with significantly
reduced GV, the so-called slow-light \cite{sj04nm,k08nphot,b08nphot}, is that both linear and
nonlinear optical effects can be dramatically enhanced in the slow-light regime
\cite{nys01prl,sjf02josab,pbo03ol,pbo04oe,vbh05n,myp06ol,cmg09nphot,rls11prb}.

One of the most important nonlinear optical process, as far as nonlinear optics applications are
concerned, is FWM. In the generic case, it consists of the combination of two photons with
frequencies, $\omega_{1}$ and $\omega_{2}$, belonging to two pump continuous-waves (CWs) or
pulses, followed by the generation of a pair of photons with frequencies $\omega_{3}$ and
$\omega_{4}$. The energy conservation requires that $\omega_{1}+\omega_{2}=\omega_{3}+\omega_{4}$.
In practice, however, an easier to implement FWM configuration is usually employed, namely
degenerate FWM. In this case one uses just one pump with frequency, $\omega_{p}$, the generated
photons belonging to a signal ($\omega_{s}$) and an idler ($\omega_{i}$) beam; in this case the
conservation of the optical energy is expressed as: $2\omega_{p}=\omega_{s}+\omega_{i}$. Among the
most important applications of degenerate FWM it is noteworthy to mention optical amplification,
wavelength generation and conversion, phase conjugation, generation of squeezed states, and
supercontinuum generation. While FWM has been investigated theoretically and experimentally in PhC
waveguides \cite{myk10oe,ssv10oe,meg10oe,jli11oe,csl12oe} and long-period Bragg waveguides
\cite{dog12oe,lzd14ol}, a comprehensive theory of FWM in silicon PhC waveguides (Si-PhCWGs), which
rigorously incorporates in a unitary way all relevant linear and nonlinear optical effects as well
as the influence of photogenerated free-carriers (FCs) on the pulse dynamics is not available yet.

In this article we introduce a rigorous theoretical model that describes FWM in Si-PhCWGs. Our
model captures the influence on the FWM process of linear optical effects, including waveguide
loss, FC dispersion (FCD) and FC absorption (FCA), nonlinear optical effects such as self- and
cross-phase modulation (SPM, XPM), two-photon absorption (TPA), and cross-absorption modulation
(XAM), as well as the mutual interaction between FCs and optical field. We also illustrate how our
model can be applied to investigate the characteristics of FWM in the slow- and fast-light
regimes, showing among other things that by incorporating the effects of FCs on the optical pulse
dynamics new physics emerge. One noteworthy example in this context is that the well-known linear
dependence of FCA on $v_{g}^{-1}$ is replaced in the slow-light regime by a $v_{g}^{-3}$ power-law
dependence.
\begin{figure}[t]
\centerline{\includegraphics[width=8.0cm]{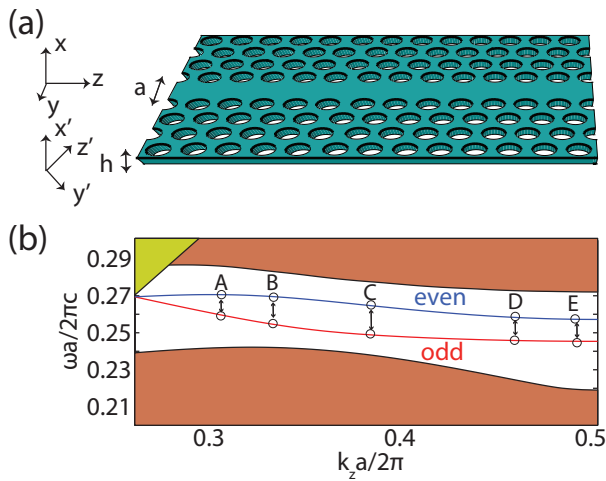}} \caption{(a) Geometry of the 1D Si-PhC
slab waveguide. The height of the slab is $h=0.6a$ and the radius of the holes is $r=0.22a$. The
primed coordinate system shows the principal axes of the Si crystal with the input facet of the
waveguide in the (1$\bar{1}$0) plane of the Si crystal lattice. (b) Projected band structure. Dark
yellow and brown areas correspond to slab leaky and guiding modes, respectively. The red and blue
curves represent the guiding modes of the 1D waveguides.} \label{fig:geom}
\end{figure}

The remaining of the paper is organized as follows. In the next section we present the optical
properties of the PhC waveguide considered in this work. Then, in Sec.~\ref{sMath}, we develop the
theory of pulsed FWM in Si-PhCWGs whereas the particular case of degenerate FWM is analyzed in
Sec.~\ref{sDFWM}. Then, in Sec.~\ref{sPMC}, we apply these theoretical tools to explore the
physical conditions in which efficient FWM can be achieved. The results are subsequently used, in
Sec.~\ref{sResults}, to study via numerical simulations the main properties of pulsed FWM in
Si-PhCWGs. We conclude our paper by summarizing in the last section the main findings of our
article and discussing some of their implications to future developments in this research area.
Finally, an averaged model that can be used in the case of broad optical pulses is presented in an
Appendix.

\begin{figure}[b]
\centerline{\includegraphics[width=8.0cm]{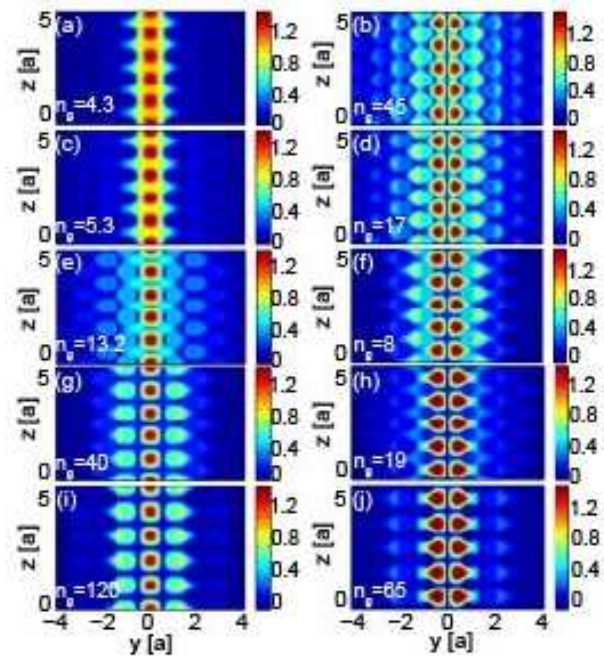}} \caption{Left (right) panels show the
amplitude of the normalized magnetic field $H_x$ of the $y$-odd ($y$-even) mode, calculated in the
plane $x=0$ for five different values of the propagation constant, $k_{z}$. From top to bottom,
the panels correspond to the Bloch modes indicated in Fig.~\ref{fig:geom}(b) by the circles
\textit{A}, \textit{B}, \textit{C}, \textit{D}, and \textit{E}, respectively.} \label{fig:modes}
\end{figure}
\section{Description of the photonic crystal waveguide}\label{sGeometry}
In this section we present the geometrical and material properties of the PhC waveguide considered
in this work, as well as the physical properties of its optical modes. Thus, our Si-PhCWG consists
of a one-dimensional (1D) waveguide formed by introducing a line defect in a two-dimensional (2D)
honeycomb-type periodic lattice of air holes in a homogeneous slab made of silicon (a so-called W1
PhC waveguide). The line defect is oriented along the $z$-axis, which is chosen to coincide with
one of the $\Gamma K$ symmetry axes of the crystal, and is created by filling in a row of holes
[see Fig.~\ref{fig:geom}(a)]. The slab height is $h=0.6a$ and the radius of the holes is
$r=0.22a$, where $a=\SI{412}{\nano\meter}$ is the lattice constant, whereas the index of
refraction of silicon is $n_{\mathrm{Si}}\equiv n=3.48$.

The defect line breaks the discrete translational symmetry of the photonic system along the
$y$-axis, so that the optical modes of the waveguide are invariant only to discrete translation
along the $z$-axis \cite{j08book}. Moreover, based on experimental considerations, we restrict our
analysis to in-plane wave propagation, namely the wave vector, $\mathbf{k}$, lies in the $x=0$
plane. The $k_{z}$ component, on the other hand, can be restricted to the first Brillouin zone,
$k_{z}\in[-\pi/a,\pi/a]$, which is an immediate consequence of the Bloch theorem. Under these
circumstances, we determined numerically the photonic band structure of the system and the guiding
optical modes of the waveguide using MPB, a freely available code based on the plane-wave
expansion (PWE) method \cite{jj01oe}. To be more specific, we used a supercell with size of $6a
\times 19 \sqrt{3}/2 a \times a$ along the $x$-, $y$-, and $z$-axis, respectively, the
corresponding step size of the computational grid being $a/60$, $a\sqrt{3}/120$, and $a/60$,
respectively. Figure~\ref{fig:geom}(b) summarizes the results of these calculations. Thus, the
waveguide has two fundamental TE-like optical guiding modes located in the band-gap of the
unperturbed PhC, one $y$-even and the other one $y$-odd.
\begin{figure}[t]
\centerline{\includegraphics[width=8.0cm]{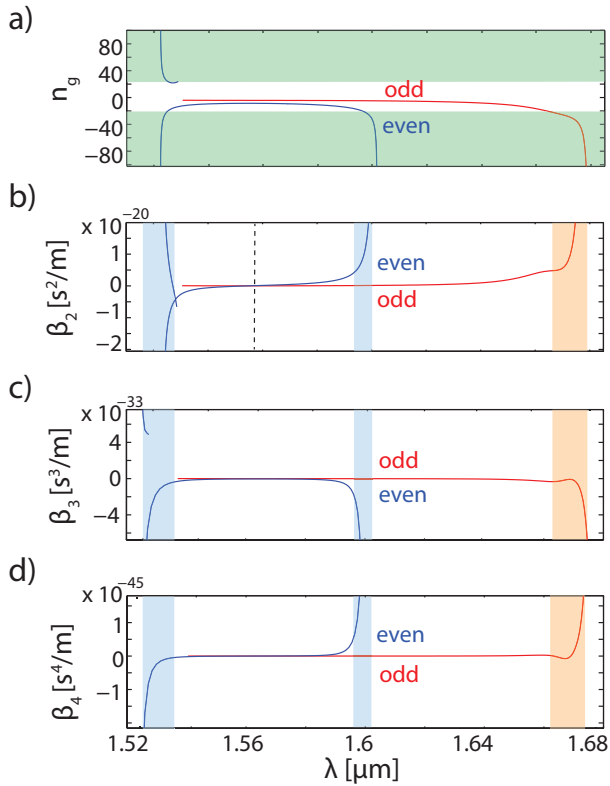}} \caption{(a), (b), (c), and (d) Frequency
dependence of waveguide dispersion coefficients $n_g$, $\beta_2$, $\beta_3$, and $\beta_4$,
respectively, determined for the even and odd modes. Light green, blue, and red shaded regions
correspond to slow-light regime, defined as $n_g>20$. The dashed vertical line in panel (b)
indicates the zero-GVD wavelength.} \label{fig:disp}
\end{figure}

In order to better understand the physical properties of the optical guiding modes, we plot in
Fig.~\ref{fig:modes} the profile of the magnetic field $H_{x}$, which is its only nonzero
component in the $x=0$ symmetry plane. These field profiles, calculated for several values of
$k_{z}$, show that although the optical field is primarily confined at the location of the defect
(waveguide), for some values of $k_{z}$ it is rather delocalized in the transverse direction. This
field delocalization effect is particularly strong in the spectral domains where the modal
dispersion curves are relatively flat, namely in the so-called slow-light regime, and increases
when the group index of the mode, defined as $n_g=c/v_g$, increases.

The dispersion effects upon pulse propagation in the waveguide are characterized by the waveguide
dispersion coefficients, defined as $\beta_n=d^nk_{z}/d\omega^n$. In particular, the first-order
dispersion coefficient is related to the pulse GV via $\beta_1=1/v_g$, whereas the second-order
dispersion coefficient, $\beta_2$, quantifies the GV dispersion (GVD) as well as pulse broadening
effects. The wavelength dependence of the first four dispersion coefficients, determined for both
guided modes, is presented in Fig.~\ref{fig:disp}, the shaded areas indicating the spectral
regions of slow-light. For the sake of clarity, we set the corresponding threshold to
$c/v_{g}=20$, that is the slow-light regime is defined by $n_g>20$. As it can be seen in
Fig.~\ref{fig:disp}, the even mode possesses two slow-light regions, one located at the band-edge
($\lambda\approx\SI{1.6}{\micro\meter}$) and the other one at $k_z\approx 0.3(2\pi/a)$, i.e.
$\lambda\approx\SI{1.52}{\micro\meter}$, whereas the odd mode contains only one such spectral
domain located at the band-edge ($\lambda\approx\SI{1.67}{\micro\meter}$). Moreover, the even mode
can have both positive and negative GVD, the zero-GVD point being at
$\lambda=\SI{1.56}{\micro\meter}$, whereas the odd mode has normal GVD ($\beta_{2}>0$) throughout.
Since usually efficient FWM can only be achieved in the anomalous GVD regime ($\beta_{2}<0$), we
will assume that the interacting pulses propagate in the even mode unless otherwise is specified.

\section{Derivation of the Mathematical Model}\label{sMath}
This section is devoted to the derivation of a system of coupled-mode equations describing the
co-propagation of a set of mutually interacting optical pulses in a Si-PhCWG, as well the
influence of photogenerated FCs on the pulse evolution. We will derive these coupled-mode
equations in the most general setting, namely the nondegenerate FWM, then show how they can be
applied to a particular case most used in practice, the so-called degenerate FWM configuration.
Our derivation follows the general approach used to develop a theoretical model for pulse
propagation in silicon waveguides with uniform cross-section \cite{cpo06jqe} and Si-PhCWGs
\cite{pmw10jstqe}.

\subsection{Optical modes of photonic crystal waveguides}
In the presence of an external perturbation described by the polarization,
$\mathbf{P}_{\mathrm{pert}}(\mathbf{r},\omega)$, the electromagnetic field of guiding modes with
frequency, $\omega$, is described by the Maxwell equations, which in the frequency domain can be
written in the following form:
\begin{subequations}\label{Me}
\begin{align}
&\nabla\times \mathbf{E}(\mathbf{r},\omega)=i\omega\mu\mathbf{H}(\mathbf{r},\omega),\label{ee} \\
&\nabla\times
\mathbf{H}(\mathbf{r},\omega)=-i\omega[\epsilon_{c}(\mathbf{r},\omega)\mathbf{E}(\mathbf{r},\omega)+\mathbf{P}_{\mathrm{pert}}(\mathbf{r},\omega)],\label{he}
\end{align}
\end{subequations}
where $\mu$ is the magnetic permeability, which in the case of silicon and other nonmagnetic
materials can be set to $\mu=\mu_{0}$, $\epsilon_{c}(\mathbf{r},\omega)$ is the dielectric
constant of the PhC, and $\mathbf{E}$ and $\mathbf{H}$ are the electric and magnetic fields,
respectively. In our case, $\mathbf{P}_{\mathrm{pert}}$ is the sum of polarizations describing the
refraction index change induced by photogenerated FCs and nonlinear (Kerr) effects.

In order to understand how the modes of the PhC waveguide are affected by external perturbations,
let us consider first the unperturbed system, that is $\mathbf{P}_{\mathrm{pert}}=0$. Thus, let us
assume that, at the frequency $\omega$, the unperturbed PhC waveguide has $M$ guiding modes. It
follows then from the Bloch theorem that the fields of these modes can be written as:
\begin{subequations}\label{modedef}
\begin{align}
&\mathbf{E}_{m\sigma}(\mathbf{r},\omega)=\mathbf{e}_{m\sigma}(\mathbf{r},\omega)e^{i\sigma\beta_{m}z},~~~m=1,2,\ldots,M, \\
&\mathbf{H}_{m\sigma}(\mathbf{r},\omega)=\mathbf{h}_{m\sigma}(\mathbf{r},\omega)e^{i\sigma\beta_{m}z},~~~m=1,2,\ldots,M,
\end{align}
\end{subequations}
where $\beta_{m}$ is the $m$th mode propagation constant and $\sigma=+$ ($\sigma=-$) denotes
forward (backward) propagating modes. Here, we consider that the harmonic time dependence of the
fields was chosen as $e^{-i\omega t}$. The mode amplitudes $\mathbf{e}_{m\sigma}$ and
$\mathbf{h}_{m\sigma}$ are periodic along the $z$-axis, with period $a$. Moreover, the forward and
backward propagating modes obey the following symmetry relations:
\begin{subequations}\label{modesymm}
\begin{align}
&\mathbf{e}_{m-}(\mathbf{r},\omega)=\mathbf{e}_{m+}^{*}(\mathbf{r},\omega), \\
&\mathbf{h}_{m-}(\mathbf{r},\omega)=-\mathbf{h}_{m+}^{*}(\mathbf{r},\omega),
\end{align}
\end{subequations}
where the symbol ``$^{*}$'' denotes complex conjugation. As such, one only has to determine either
the forward or the backward propagating modes.

The guiding modes can be orthogonalized, the most commonly used normalization convention being
\begin{equation}\label{modeorth}
\frac{1}{4}\int_{S}\left(\mathbf{e}_{m\sigma}\times\mathbf{h}_{m'\sigma'}^{*}+\mathbf{e}_{m'\sigma'}^{*}\times\mathbf{h}_{m\sigma}\right)\cdot\hat{\mathbf{z}}dS=\sigma
P_{m}\delta_{\sigma\sigma'}\delta_{mm'},
\end{equation}
where $P_{m}$ is the power carried by the $m$th mode. This mode power is related to the mode
energy contained in one unit cell of the PhC waveguide, $W_{m}$, via the relation:
\begin{equation}\label{PEnDens}
P_{m}=\frac{W_{m}^{\mathrm{el}}+W_{m}^{\mathrm{mag}}}{a}v_{g} =
\frac{2W_{m}^{\mathrm{el}}}{a}v_{g} = \frac{2W_{m}^{\mathrm{mag}}}{a}v_{g},
\end{equation}
where
\begin{subequations}\label{energEH}
\begin{align}
\label{energE}
&W_{m}^{\mathrm{el}} = \frac{1}{4}\int_{V_{\mathrm{cell}}}\frac{\partial}{\partial\omega}(\omega\epsilon_{c})\vert\mathbf{e}_{m\sigma}(\mathbf{r},\omega)\vert^{2}dV, \\
\label{energH} &W_{m}^{\mathrm{mag}} =
\frac{1}{4}\int_{V_{\mathrm{cell}}}\mu_{0}\vert\mathbf{h}_{m\sigma}(\mathbf{r},\omega)\vert^{2}dV,
\end{align}
\end{subequations}
are the electric and magnetic energy of the mode, respectively, and $V_{\mathrm{cell}}$ is the
volume of the unit cell. Note that in Eq.~\eqref{PEnDens} we used the fact that the mode contains
equal amounts of electric and magnetic energy.

It should be stressed that the waveguide modes defined by Eqs.~\eqref{modedef} are exact solutions
of the Maxwell equations \eqref{Me} with $\mathbf{P}_{\mathrm{pert}}=0$, and thus they should not
be confused with the so-called local modes of the waveguide. The latter modes correspond to
waveguides whose optical properties vary adiabatically with $z$, on a scale comparable to the
wavelength and have been used to describe, \textit{e.g.}, wave propagation in tapered waveguides
\cite{s70tmtt} or pulse propagation in 1D long-period Bragg gratings \cite{sse02jmo}.

\subsection{Perturbations of the photonic crystal waveguide}
Due to the photogeneration of FCs and nonlinear optical effects, the dielectric constant of
Si-PhCWGs undergoes a certain local variation, $\delta\epsilon(\mathbf{r})$, upon the propagation
of optical pulses in the waveguide. The corresponding perturbation polarization,
$\mathbf{P}_{\mathrm{pert}}$ in Eq.~\eqref{he}, can be divided in two components according to the
physical effects they describe: the linear change of the dielectric constant \textit{via}
generation of FCs and the nonlinearly induced variation of the index of refraction.

Assuming an instantaneous response of the medium, the linear contribution to
$\mathbf{P}_{\mathrm{pert}}$, $\delta \mathbf{P}_{\mathrm{lin}}(\mathbf{r},t)$, is written as:
\begin{equation}\label{linp}
\delta
\mathbf{P}_{\mathrm{lin}}(\mathbf{r},t)=\left[\delta\epsilon_{\mathrm{fc}}(\mathbf{r})+\delta\epsilon_{\mathrm{loss}}(\mathbf{r})\right]\mathbf{E}(\mathbf{r},t),
\end{equation}
where \cite{cpo06jqe}:
\begin{subequations}\label{epslin}
\begin{align}
\label{epslinfc}&\delta \epsilon_{\mathrm{fc}}(\mathbf{r}) = \left(2\epsilon_{0}n\delta
n_{\mathrm{fc}} + i\frac{\displaystyle \epsilon_{0}cn}{\displaystyle
\omega}\alpha_{\mathrm{fc}}\right)\Sigma(\mathbf{r}),\\
\label{epslinfc}&\delta\epsilon_{\mathrm{loss}}(\mathbf{r}) = i\frac{\displaystyle
\epsilon_{0}cn}{\displaystyle \omega}\alpha_{\mathrm{in}}\Sigma(\mathbf{r}).
\end{align}
\end{subequations}
Here, $\alpha_{\mathrm{in}}$ is the intrinsic loss coefficient of the waveguide and
$\Sigma(\mathbf{r})$ is the characteristic function of the domain where FCs can be generated,
namely $\Sigma=1$ in the domain occupied by Si and $\Sigma=0$ otherwise. Based on the Drude model,
the FC-induced change of the index of refraction, $\delta n_{\mathrm{fc}}$, and FC losses,
$\alpha_{\mathrm{fc}}$, are given by \cite{sb87jqe}:
\begin{subequations}
\label{nepsFC}
\begin{align}
\label{nFC}& \delta n_{\mathrm{fc}} = -\frac{\displaystyle e^{2}}{\displaystyle
2\epsilon_{0}n\omega^{2}}
\left ( \frac{N_{e}}{m_{ce}} + \frac{N_{h}^{0.8}}{m_{ch}} \right ), \\
\label{epsFC}& \alpha_{\mathrm{fc}} = \frac{\displaystyle e^{3}}{\displaystyle
\epsilon_{0}cn\omega^{2}} \left ( \frac{N_{e}}{\mu_{e}m_{ce}^{2}} +
\frac{N_{h}}{\mu_{h}m_{ch}^{2}} \right ).
\end{align}
\end{subequations}
Here, $e$ is the charge of the electron, $\mu_{e}$ ($\mu_{h}$) is the electron (hole) mobility,
$m_{ce}=0.26m_{0}$ ($m_{ch}=0.39m_{0}$) is the conductivity effective mass of the electrons
(holes), with $m_{0}$ the mass of the electron, and $N_{e}$ ($N_{h}$) is the induced variation of
the electrons (holes) density (in what follows, we assume that $N_{e}=N_{h}\equiv N$).

The nonlinear contribution to $\mathbf{P}_{\mathrm{pert}}$, $\delta
\mathbf{P}_{\mathrm{nl}}(\mathbf{r},t)$, is described by a third-order nonlinear susceptibility,
$\hat{\chi}^{(3)}(\mathbf{r})$, and can be written as:
\begin{equation}\label{Kerr}
\delta \mathbf{P}_{\mathrm{nl}}(\mathbf{r},t)=\epsilon_{0}\hat{\chi}^{(3)}(\mathbf{r})
\vdots\mathbf{E}(\mathbf{r},t)\mathbf{E}(\mathbf{r},t)\mathbf{E}(\mathbf{r},t).
\end{equation}
The real part of the susceptibility $\hat{\chi}^{(3)}$ describes parametric optical processes such
as SPM, XPM, and FWM, while the imaginary part of $\hat{\chi}^{(3)}$ corresponds to TPA and XAM.
Note that in this study we neglect the stimulated Raman scattering effect as it is assumed that
the frequencies of the interacting pulses do not satisfy the condition required for an efficient,
resonant Raman interaction.

Since silicon belongs to the crystallographic point group $m3m$ the susceptibility tensor
$\mathbb{\hat{\chi}}^{(3)}$ has 21 nonzero elements, of which only 4 are independent, namely,
$\mathbb{\chi}_{1111}$, $\mathbb{\chi}_{1122}$, $\mathbb{\chi}_{1212}$, and $\mathbb{\chi}_{1221}$
\cite{b08book}. In addition, the frequency dispersion of the nonlinear susceptibility can be
neglected as we consider optical pulses with duration of just a few picoseconds or larger. As a
consequence, the Kleinman symmetry relations imply that
$\mathbb{\chi}_{1122}=\mathbb{\chi}_{1212}=\mathbb{\chi}_{1221}$. Moreover, experimental studies
have shown that $\hat{\chi}^{(3)}_{1111} = 2.36\hat{\chi}^{(3)}_{1122}$ \cite{zlp07apl} within a
broad frequency range. Therefore, the nonlinear optical effects considered here can be described
by only one element of the tensor $\hat{\chi}^{(3)}$.

Because of fabrication considerations, in many instances the waveguide is not aligned with any of
the crystal principal axes and as such these axes are different from the coordinate axes in which
the optical modes are calculated. Therefore, one has to transform the tensor $\hat{\chi}^{(3)}$
from the crystal principal axes into the coordinate system in which the optical modes are
calculated \cite{cpo06jqe},
\begin{equation}\label{rotChi}
\hat{\chi}^{(3)}_{ijkl} =
\hat{R}_{i\alpha}\hat{R}_{j\beta}\hat{R}_{k\gamma}\hat{R}_{l\delta}\hat{\chi}^{\prime(3)}_{\alpha\beta\gamma\delta},
\end{equation}
where $\hat{\chi}^{\prime(3)}$ is the nonlinear susceptibility in the crystal principal axes and
$\hat{R}$ is the rotation matrix that transforms one coordinate system into the other. In our
case, $\hat{R}$ is the matrix describing a rotation with $\pi/4$ around the $x$-axis (see
Fig.~\ref{fig:geom}).

\subsection{Coupled-mode equations for the optical field}
In order to derive the system of coupled-mode equations describing pulsed FWM in Si-PhCWGs we
employ the conjugated form of the Lorentz reciprocity theorem
\cite{cpo06jqe,mpw03pre,ks07jqe,s83book}. To this end, let us consider two solutions of the
Maxwell equations \eqref{Me},
$[\mathbf{E}_{a}(\mathbf{r},\omega_{a}),\mathbf{H}_{a}(\mathbf{r},\omega_{a})]$ and
$[\mathbf{E}_{b}(\mathbf{r},\omega_{b}),\mathbf{H}_{b}(\mathbf{r},\omega_{b})]$, which correspond
to two different spatial distribution of the dielectric constant,
$\epsilon_{a}(\mathbf{r},\omega_{a})$ and $\epsilon_{b}(\mathbf{r},\omega_{b})$, respectively. If
we insert the vector $\mathbf{F}$, defined as $\mathbf{F} = \mathbf{E}_{b} \times
\mathbf{H}_{a}^{*} + \mathbf{E}_{a}^{*} \times \mathbf{H}_{b}$, in the integral identity:
\begin{equation}\label{Lorentz}
\int_{S} \nabla \cdot \mathbf{F} dS = \frac{\displaystyle \partial}{\displaystyle \partial z}
\int_{S} \mathbf{F} \cdot \hat{\mathbf{z}} dS + \oint_{\partial S}\mathbf{F}\cdot\mathbf{n} dl,
\end{equation}
where $S$ is the transverse section at position, $z$, and $\partial S$ is the boundary of $S$, and
use the Maxwell equations, we arrive at the following relation:
\begin{align}\label{Lorentzred}
\frac{\displaystyle \partial}{\displaystyle \partial z} &\int_{S} \mathbf{F} \cdot
\hat{\mathbf{z}}dS = i\mu_{0}(\omega_{b}-\omega_{a})\int_{S} \mathbf{H}_{a}^{*} \cdot \mathbf{H}_{b} dS \notag \\
&+i\int_{S}(\omega_{b}\epsilon_{b}-\omega_{a}\epsilon_{a}) \mathbf{E}_{a}^{*} \cdot \mathbf{E}_{b}
dS-\oint_{\partial S}\mathbf{F}\cdot\mathbf{n} dl.
\end{align}

Let us consider now a nondegenerate FWM process in which two pulses at carrier frequencies
$\bar{\omega}_{1}$ and $\bar{\omega}_{2}$ interact and generate two optical pulses at carrier
frequencies $\bar{\omega}_{3}$ and $\bar{\omega}_{4}$, with the energy conservation expressed as
$\bar{\omega}_{1}+\bar{\omega}_{2}=\bar{\omega}_{3}+\bar{\omega}_{4}$. Then, in the Lorentz
reciprocity theorem given by Eq.~\eqref{Lorentzred} we choose as the first set of fields a mode of
the unperturbed waveguide ($\mathbf{P}_{\mathrm{pert}}=0$), which corresponds to the frequency
$\omega_{a}=\bar{\omega}_{i}$, where $\bar{\omega}_{i}$ is one of the carrier frequencies
$\bar{\omega}_{1}$, $\bar{\omega}_{2}$, $\bar{\omega}_{3}$, or $\bar{\omega}_{4}$:
\begin{subequations}\label{field1}
\begin{align}
&\mathbf{E}_{a}(\mathbf{r},\bar{\omega}_{i})=\frac{\mathbf{e}_{n_{i}\rho_{i}}(\mathbf{r},\bar{\omega}_{i})}{\sqrt{\bar{P}_{n_{i}}}}e^{i\rho_{i}\bar{\beta}_{n_{i}}z}, \\
&\mathbf{H}_{a}(\mathbf{r},\bar{\omega}_{i})=\frac{\mathbf{h}_{n_{i}\rho_{i}}(\mathbf{r},\bar{\omega}_{i})}{\sqrt{\bar{P}_{n_{i}}}}e^{i\rho_{i}\bar{\beta}_{n_{i}}z},
\end{align}
\end{subequations}
where $\rho_{i}=\pm1$ and $n_{i}$ is an integer, $1\leq n_{i}\leq N_{i}$, $i=1,\ldots,4$, with
$N_{i}$ being the number of guiding modes at the frequency $\bar{\omega}_{i}$. In
Eqs.~\eqref{field1}, and in what follows, a bar over a symbol means that the corresponding
quantity is evaluated at one of the carrier frequencies.

As the second set of fields we take those that propagate in the perturbed waveguide, at the
frequency $\omega_{b}=\omega$. These fields are written as a series expansion of the guiding modes
at frequencies $\bar{\omega}_{i}$, $i=1,\ldots,4$, thus neglecting the frequency dispersion of the
guiding modes and the radiative modes that might exist at the frequency $\omega$. This
approximation is valid as long as all interacting optical pulses have narrow spectra centered at
the corresponding carrier frequencies, that is the physical situation considered in this work. In
particular, this modal expansion becomes less accurate when any of the pulses propagates in the
slow-light regime, as generally the smaller the GV of a mode is the larger its frequency
dispersion is. Thus, the second set of fields are expanded as:
\begin{subequations}\label{field2}
\begin{align}
&\mathbf{E}_{b}(\mathbf{r},\omega)=\sum_{j=1}^{4}\sum_{m_{j}\sigma_{j}}a_{m_{j}\sigma_{j}}^{(j)}(z,\omega)\frac{\mathbf{e}_{m_{j}\sigma_{j}}(\mathbf{r},\bar{\omega}_{j})}{\sqrt{\bar{P}_{m_{j}}}}e^{i\sigma_{j}\bar{\beta}_{m_{j}}z}, \\
&\mathbf{H}_{b}(\mathbf{r},\omega)=\sum_{j=1}^{4}\sum_{m_{j}\sigma_{j}}a_{m_{j}\sigma_{j}}^{(j)}(z,\omega)\frac{\mathbf{h}_{m_{j}\sigma_{j}}(\mathbf{r},\bar{\omega}_{j})}{\sqrt{\bar{P}_{m_{j}}}}e^{i\sigma_{j}\bar{\beta}_{m_{j}}z}.
\end{align}
\end{subequations}
With the fields normalization used in Eqs.~\eqref{field2}, the mode amplitudes
$a_{m_{i}\sigma_{i}}^{(i)}(z,\omega)$, $i=1,\ldots,4$, are measured in units of
$\sqrt{\mathrm{W}}$. Note that since the optical pulses are assumed to be spectrally narrow, the
mode amplitudes $a_{m_{i}\sigma_{i}}^{(i)}(z,\omega)$ have negligible values except when the
frequency $\omega$ lies in a narrow spectral domain centered at the carrier frequency,
$\bar{\omega}_{i}$.

The dielectric constant in the two cases is
$\epsilon_{a}=\bar{\epsilon}_{c}(\mathbf{r},\bar{\omega}_{i})$ and
$\epsilon_{b}=\epsilon_{c}(\mathbf{r},\omega)+\delta\epsilon(\mathbf{r},\omega)$, where
$\epsilon_{c}(\mathbf{r},\omega)$ is the dielectric constant of the unperturbed PhC. If the
material dispersion is neglected,
$\epsilon_{c}(\mathbf{r},\omega)=\epsilon_{c}(\mathbf{r},\bar{\omega}_{i})=\bar{\epsilon}_{c}(\mathbf{r})$.
Inserting the fields given by Eqs.~\eqref{field1} and Eqs.~\eqref{field2} in
Eq.~\eqref{Lorentzred}, and neglecting the line integral in Eq.~\eqref{Lorentzred}, which cancels
for exponentially decaying guiding modes, one obtains the following set of coupled equations:
\begin{align}\label{cmefreq}
\rho_{i}&\frac{\displaystyle \partial a_{n_{i}\rho_{i}}^{(i)}(z)}{\displaystyle \partial z} +
\sum\limits_{\scriptsize{\begin{array}{c}
        j=1 \\
        j\neq i
      \end{array}}}^{4}\sum_{m_{j}\sigma_{j}}C_{n_{i}\rho_{i},m_{j}\sigma_{j}}^{ij}(z) \notag \\&
      \times\left[\frac{\displaystyle \partial a_{m_{j}\sigma_{j}}^{(j)}(z)}{\displaystyle \partial z}
      +i(\sigma_{j}\bar{\beta}_{m_{j}}-\rho_{i}\bar{\beta}_{n_{i}})a_{m_{j}\sigma_{j}}^{(j)}(z)\right] \notag \\
      &= B_{n_{i}\rho_{i}}^{i}a_{n_{i}\rho_{i}}^{(i)}(z) +
{\sum_{jm_{j}\sigma_{j}}}^{\prime}D_{n_{i}\rho_{i},m_{j}\sigma_{j}}^{ij}a_{m_{j}\sigma_{j}}^{(j)}(z) \notag \\
&+\frac{i\omega
e^{-i\rho_{i}\bar{\beta}_{n_{i}}z}}{4\sqrt{\bar{P}_{n_{i}}}}\int_{S}\bar{\mathbf{e}}_{n_{i}\rho_{i}}^{*}
\cdot \mathbf{P}_{\mathrm{pert}}(\mathbf{r},\omega) dS,~~~i=1,\ldots,4,
\end{align}
where
\begin{subequations}\label{dispcoeffgen}
\begin{align}
\label{dispcoeffgenC}
C_{n_{i}\rho_{i},m_{j}\sigma_{j}}^{ij}&(z)=\frac{e^{i(\sigma_{j}\bar{\beta}_{m_{j}}-\rho_{i}\bar{\beta}_{n_{i}})z}}{4\sqrt{\bar{P}_{n_{i}}\bar{P}_{m_{j}}}}\notag\\
&\times\int_{S}\left(\bar{\mathbf{e}}_{m_{j}\sigma_{j}}\times\bar{\mathbf{h}}_{n_{i}\rho_{i}}^{*}+\bar{\mathbf{e}}_{n_{i}\rho_{i}}^{*}\times\bar{\mathbf{h}}_{m_{j}\sigma_{j}}\right)\cdot\hat{\mathbf{z}}dS,
\\
\label{dispcoeffgenB}
B_{n_{i}\rho_{i}}^{i}&=\frac{i}{4\bar{P}_{n_{i}}}\int_{S}\left[\mu_{0}(\omega-\bar{\omega}_{i})\vert\bar{\mathbf{h}}_{n_{i}\rho_{i}}\vert^{2}\right.
\notag\\&\left.+(\omega\epsilon_{c}-\bar{\epsilon}_{c}\bar{\omega}_{i})\vert\bar{\mathbf{e}}_{n_{i}\rho_{i}}\vert^{2}\right]dS, \\
\label{dispcoeffgenD}
D_{n_{i}\rho_{i},m_{j}\sigma_{j}}^{ij}&(z)=\frac{ie^{i(\sigma_{j}\bar{\beta}_{m_{j}}-\rho_{i}\bar{\beta}_{n_{i}})z}}{4\sqrt{\bar{P}_{n_{i}}\bar{P}_{m_{j}}}}
\int_{S}\left[\mu_{0}(\omega-\bar{\omega}_{i})\right.\notag\\&\left.\times\bar{\mathbf{h}}_{m_{j}\sigma_{j}}\cdot\bar{\mathbf{h}}_{n_{i}\rho_{i}}^{*}
+(\omega\epsilon_{c}-\bar{\epsilon}_{c}\bar{\omega}_{i})\bar{\mathbf{e}}_{m_{j}\sigma_{j}}\cdot\bar{\mathbf{e}}_{n_{i}\rho_{i}}^{*}\right]dS.
\end{align}
\end{subequations}
In Eq.~\eqref{cmefreq} and what follows a prime symbol to a sum means that the summation is taken
over all modes, except that with $j=i$, $m_{j}=n_{i}$, and $\sigma_{j}=\rho_{i}$. Moreover, in
deriving the l.h.s. of Eq.~\eqref{cmefreq} we used the orthogonality relation given by
Eq.~\eqref{modeorth}.

The time-dependent fields are obtained by integrating over all frequency components contained in
the spectra of the system of interacting optical pulses:
\begin{subequations}\label{fieldtime}
\begin{align}
\label{fieldtimeE}
\mathbf{E}&(\mathbf{r},t)=\frac{1}{2}\int_{0}^{\infty}\sum_{j=1}^{4}\sum_{m_{j}\sigma_{j}}a_{m_{j}\sigma_{j}}^{(j)}(z,\omega)\frac{\mathbf{e}_{m_{j}\sigma_{j}}(\mathbf{r},\bar{\omega}_{j})}{\sqrt{\bar{P}_{m_{j}}}}
 \notag \\ &\times e^{i(\sigma_{j}\bar{\beta}_{m_{j}}z-\omega t)} d\omega+c.c.\equiv\frac{1}{2}\left[\mathbf{E}^{(+)}(\mathbf{r},t)+\mathbf{E}^{(-)}(\mathbf{r},t)\right], \\
\label{fieldtimeH}
\mathbf{H}&(\mathbf{r},t)=\frac{1}{2}\int_{0}^{\infty}\sum_{j=1}^{4}\sum_{m_{j}\sigma_{j}}a_{m_{j}\sigma_{j}}^{(j)}(z,\omega)\frac{\mathbf{h}_{m_{j}\sigma_{j}}(\mathbf{r},\bar{\omega}_{j})}{\sqrt{\bar{P}_{m_{j}}}}
 \notag
\\ &\times e^{i(\sigma_{j}\bar{\beta}_{m_{j}}z-\omega t)}d\omega+c.c.\equiv\frac{1}{2}\left[\mathbf{H}^{(+)}(\mathbf{r},t)+\mathbf{H}^{(-)}(\mathbf{r},t)\right],
\end{align}
\end{subequations}
where $\mathbf{E}^{(+)}(\mathbf{r},t)$, $\mathbf{H}^{(+)}(\mathbf{r},t)$ and
$\mathbf{E}^{(-)}(\mathbf{r},t)$, $\mathbf{H}^{(-)}(\mathbf{r},t)$ are the positive and negative
frequency parts of the spectrum, respectively.

Let us now introduce the envelopes of the interacting pulses in the time domain,
$A_{n_{i}\rho_{i}}^{(i)}(z,t)$, defined as the integral of the mode amplitudes taken over the part
of the spectrum that contains only positive frequencies,
\begin{align}\label{env}
A_{n_{i}\rho_{i}}^{(i)}(z,t) =
\int_{0}^{\infty}a_{n_{i}\rho_{i}}^{(i)}(z,\omega)e^{-i(\omega-\bar{\omega}_{i})t}d\omega.
\end{align}
With this definition, the time-dependent fields given in Eqs.~\eqref{fieldtime} become:
\begin{subequations}\label{fieldtimesimpl}
\begin{align}
\mathbf{E}(\mathbf{r},t)=&\frac{1}{2}\sum_{j=1}^{4}\sum_{m_{j}\sigma_{j}}A_{m_{j}\sigma_{j}}^{(j)}(z,t)\notag \\
&\times\frac{\bar{\mathbf{e}}_{m_{j}\sigma_{j}}(\mathbf{r},\bar{\omega}_{j})}{\sqrt{\bar{P}_{m_{j}}}}e^{i(\sigma_{j}\bar{\beta}_{m_{j}}z-\bar{\omega}_{j} t)}+c.c., \\
\mathbf{H}(\mathbf{r},t)=&\frac{1}{2}\sum_{j=1}^{4}\sum_{m_{j}\sigma_{j}}A_{m_{j}\sigma_{j}}^{(j)}(z,t)\notag \\
&\times\frac{\bar{\mathbf{h}}_{m_{j}\sigma_{j}}(\mathbf{r},\bar{\omega}_{j})}{\sqrt{\bar{P}_{m_{j}}}}e^{i(\sigma_{j}\bar{\beta}_{m_{j}}z-\bar{\omega}_{j}
t)}+c.c..
\end{align}
\end{subequations}
Following the same approach, the time-dependent polarization, too, can be decomposed in two
components, which contain positive and negative frequencies, that is, it can be written as:
\begin{align}\label{poltime}
\mathbf{P}_{\mathrm{pert}}(\mathbf{r},t)&=\frac{1}{2}\int_{0}^{\infty}\mathbf{P}_{\mathrm{pert}}(\mathbf{r},\omega)e^{-i\omega
t}d\omega +c.c.\notag \\
&\equiv\frac{1}{2}\left[\mathbf{P}_{\mathrm{pert}}^{(+)}(\mathbf{r},t)+\mathbf{P}_{\mathrm{pert}}^{(-)}(\mathbf{r},t)\right].
\end{align}

The next step of our derivation is to Fourier transform Eq.~\eqref{cmefreq} in the time domain. To
this end, we first expand the coefficients $B_{n_{i}\rho_{i}}^{i}$ and
$D_{n_{i}\rho_{i},m_{j}\sigma_{j}}^{ij}$ in Taylor series, around the carrier frequency
$\bar{\omega}_{i}$ [note that according to Eq.~\eqref{dispcoeffgenC},
$C_{n_{i}\rho_{i},m_{j}\sigma_{j}}^{ij}$ is frequency independent]:
\begin{subequations}\label{dispcoeffTs}
\begin{align}
\label{dispcoeffTsB}
B_{n_{i}\rho_{i}}^{i}=\sum_{q\geq1}&\frac{(\Delta\omega_{i})^{q}}{q!}\left.\frac{\partial^{q}B_{n_{i}\rho_{i}}^{i}}{\partial\omega^{q}}\right|_{\omega=\bar{\omega}_{i}}
\equiv \sum_{q\geq1}\frac{i\beta^{(q)i}_{n_{i}\rho_{i}}}{q!}(\Delta\omega_{i})^{q}, \\
\label{dispcoeffTsD}
D_{n_{i}\rho_{i},m_{j}\sigma_{j}}^{ij}&=\sum_{q\geq1}\frac{(\Delta\omega_{i})^{q}}{q!}\left.\frac{\partial^{q}D_{n_{i}\rho_{i},m_{j}\sigma_{j}}^{ij}}{\partial\omega^{q}}\right|_{\omega=\bar{\omega}_{i}}\nonumber
\\ &\equiv \sum_{q\geq1}\frac{i\beta^{(q)ij}_{n_{i}\rho_{i},m_{j}\sigma_{j}}}{q!}(\Delta\omega_{i})^{q},
\end{align}
\end{subequations}
where $\Delta\omega_{i}=\omega-\bar{\omega}_{i}$, $i=1,\ldots,4$. Combining
Eqs.~\eqref{dispcoeffTsB}, \eqref{dispcoeffgenB}, \eqref{PEnDens}, and \eqref{energEH} leads to
the following expression for the dispersion coefficients, $\beta^{(q)i}_{n_{i}\rho_{i}}$:
\begin{subequations}\label{dispcoeffZdep}
\begin{align}
&\beta^{(1)i}_{n_{i}\rho_{i}}(z) = \frac{\delta^{i}_{n_{i}}(z)}{v_{g,n_{i}}^{i}}, \\
&\beta^{(n)i}_{n_{i}\rho_{i}}(z) =
\delta^{i}_{n_{i}}(z)\frac{\partial^{n-1}}{\partial\omega^{n-1}}\left(\frac{1}{v_{g,n_{i}}^{i}}\right),~n\geq2,
\end{align}
\end{subequations}
where
\begin{align}\label{delDef}
\delta^{i}_{n_{i}}(z)=&\frac{a}{4W_{m_{i}}}\int_{S}\left[\mu_{0}\vert\mathbf{h}_{n_{i}\rho_{i}}(\mathbf{r},\bar{\omega}_{i})\vert^{2}
\right. \notag \\
&+\left.\frac{\partial}{\partial\omega}(\omega\epsilon_{c})\vert\mathbf{e}_{n_{i}\rho_{i}}(\mathbf{r},\bar{\omega}_{i})\vert^{2}\right]dS.
\end{align}

It can be easily seen from this equation that the average of $\delta^{i}_{n_{i}}(z)$ over one
lattice cell of the PhC waveguide is equal to 1, \textit{i.e.},
\begin{equation}\label{delAv}
\tilde{\delta}^{i}_{n_{i}}\equiv\frac{1}{a}\int_{z}^{z+a}\delta^{i}_{n_{i}}(z')dz'=1.
\end{equation}
Here and in what follows the tilde symbol indicates that the corresponding physical quantity has
been averaged over a lattice cell of the waveguide. With this notation, Eqs.~\eqref{dispcoeffZdep}
become:
\begin{subequations}\label{dispcoeffZdepAv}
\begin{align}
&\tilde{\beta}^{(1)i}_{n_{i}\rho_{i}} \equiv \beta^{i}_{1,n_{i}} = \frac{1}{v_{g,n_{i}}^{i}}, \\
&\tilde{\beta}^{(n)i}_{n_{i}\rho_{i}}  \equiv \beta^{i}_{n,n_{i}} = \frac{\partial^{n-1}
\beta^{i}_{1,n_{i}}}{\partial\omega^{n-1}},~n\geq2.
\end{align}
\end{subequations}
These relations show that $\tilde{\beta}^{(n)i}_{n_{i}\rho_{i}} = \beta^{i}_{n,n_{i}}$ is the
$n$th order dispersion coefficient of the waveguide mode characterized by the parameters
$\{n_{i},\rho_{i}\}$, evaluated at $\omega=\bar{\omega}_{i}$.

We now multiply Eq.~\eqref{cmefreq} by $e^{-i(\omega-\bar{\omega}_{i})t}$ and integrate over the
positive-frequency domain. These simple calculations lead to the time-domain coupled-mode
equations for the field envelopes, $A_{n_{i}\rho_{i}}^{(i)}(z,t)$:
\begin{align}\label{cmetime}
&\rho_{i}\frac{\displaystyle \partial A_{n_{i}\rho_{i}}^{(i)}}{\displaystyle \partial z} +
\sum\limits_{\scriptsize{\begin{array}{c}
        j=1 \\
        j\neq i
      \end{array}}}^{4}\sum_{m_{j}\sigma_{j}}C_{n_{i}\rho_{i},m_{j}\sigma_{j}}^{ij}e^{-i(\bar{\omega}_{j}-\bar{\omega}_{i})t}\left[\frac{\displaystyle \partial A_{m_{j}\sigma_{j}}^{(j)}}{\displaystyle \partial z}\right. \notag \\
&\left.+i(\sigma_{j}\bar{\beta}_{m_{j}}-\rho_{i}\bar{\beta}_{n_{i}})A_{m_{j}\sigma_{j}}^{(j)}\right]= i\sum_{q\geq1}\frac{\beta^{(q)i}_{n_{i}\rho_{i}}}{q!}\left(i\frac{\partial}{\partial t}\right)^{q}A_{n_{i}\rho_{i}}^{(i)} \notag \\
&+i\sum_{q\geq1}{\sum_{jm_{j}\sigma_{j}}}^{\prime}\frac{\beta^{(q)ij}_{n_{i}\rho_{i},m_{j}\sigma_{j}}}{q!}e^{-i(\bar{\omega}_{j}-\bar{\omega}_{i})t}\left(i\frac{\partial}{\partial t}\right)^{q}A_{n_{j}\rho_{j}}^{(j)}\notag\\
&+\frac{i\bar{\omega}_{i}e^{-i(\rho_{i}\bar{\beta}_{n_{i}}z-\bar{\omega}_{i}t)}}{4\sqrt{\bar{P}_{n_{i}}}}\int_{S}\bar{\mathbf{e}}_{n_{i}\rho_{i}}^{*}
\cdot \mathbf{P}_{\mathrm{pert}}^{(+)}(\mathbf{r},t) dS,~i=1,\ldots,4,
\end{align}

The temporal width of the pulses considered in this analysis is much smaller as compared to the
nonlinear electronic response time of silicon and therefore the latter can be approximated to be
instantaneous. In addition, we assume that the spectra of the interacting pulses are narrow and do
not overlap. Under these circumstances, the optical pulses can be viewed as quasi-monochromatic
waves and their nonlinear interactions can be treated in the adiabatic limit. Separating the
nonlinear optical effects contributing to the nonlinear polarization, one can express in the time
domain this polarization as \cite{bc91book}:
\begin{widetext}
\begin{align}\label{polnltime}
&\delta\mathbf{P}_{\mathrm{nl},\bar{\omega}_{i}}(\mathbf{r},t)=\frac{3}{4}\sum_{m_{i}\sigma_{i}}\epsilon_{0}\hat{\chi}^{(3)}(\bar{\omega}_{i},-\bar{\omega}_{i},\bar{\omega}_{i})\vdots
\bar{\mathbf{e}}_{m_{i}\sigma_{i}}(\mathbf{r},\bar{\omega}_{j})\bar{\mathbf{e}}_{m_{i}\sigma_{i}}^{*}(\mathbf{r},\bar{\omega}_{j})
\bar{\mathbf{e}}_{m_{i}\sigma_{i}}(\mathbf{r},\bar{\omega}_{j})\vert
A_{m_{i}\sigma_{i}}^{(i)}{\vert}^{2}A_{m_{i}\sigma_{i}}^{(i)}\frac{e^{i\sigma_{i}\bar{\beta}_{m_{i}}z}}{\bar{P}_{m_{i}}\sqrt{\bar{P}_{m_{i}}}} \notag \\
&~+\frac{3}{2}\sum_{m_{i}\sigma_{i}}\sum\limits_{\scriptsize{\begin{array}{c}
        (p_{i}\varrho_{i})\neq(m_{i}\sigma_{i}) \\
        p_{i}>m_{i}
      \end{array}}}\epsilon_{0}\hat{\chi}^{(3)}(\bar{\omega}_{i},-\bar{\omega}_{i},\bar{\omega}_{i})\vdots
\bar{\mathbf{e}}_{p_{i}\varrho_{i}}(\mathbf{r},\bar{\omega}_{i})\bar{\mathbf{e}}_{p_{i}\varrho_{i}}^{*}(\mathbf{r},\bar{\omega}_{i})
\bar{\mathbf{e}}_{m_{i}\sigma_{i}}(\mathbf{r},\bar{\omega}_{i})\vert
A_{p_{i}\varrho_{i}}^{(i)}{\vert}^{2}A_{m_{i}\sigma_{i}}^{(i)}\frac{e^{i\sigma_{i}\bar{\beta}_{m_{i}}z}}{\bar{P}_{p_{i}}\sqrt{\bar{P}_{m_{i}}}} \notag \\
&~+\frac{3}{2}\sum\limits_{\scriptsize{\begin{array}{c}
        j=1 \\
        j\neq i
      \end{array}}}^{4}\sum\limits_{\scriptsize{\begin{array}{c}
        m_{i}\sigma_{i} \\
        p_{j}\varrho_{j}
      \end{array}}}\epsilon_{0}\hat{\chi}^{(3)}(\bar{\omega}_{j},-\bar{\omega}_{j},\bar{\omega}_{i})\vdots
\bar{\mathbf{e}}_{p_{j}\varrho_{j}}(\mathbf{r},\bar{\omega}_{j})\bar{\mathbf{e}}_{p_{j}\varrho_{j}}^{*}(\mathbf{r},\bar{\omega}_{j})
\bar{\mathbf{e}}_{m_{i}\sigma_{i}}(\mathbf{r},\bar{\omega}_{i})\vert
A_{p_{j}\varrho_{j}}^{(j)}{\vert}^{2}A_{m_{i}\sigma_{i}}^{(i)}\frac{e^{i\sigma_{i}\bar{\beta}_{m_{i}}z}}{\bar{P}_{p_{j}}\sqrt{\bar{P}_{m_{i}}}} \notag \\
&~+\frac{3}{2}\sum\limits_{\scriptsize{\begin{array}{c}
        p_{j}q_{k}m_{l} \\
        \varrho_{j}\tau_{k}\sigma_{l}
      \end{array}}}\epsilon_{0}\hat{\chi}^{(3)}(\bar{\omega}_{j},-\bar{\omega}_{k},\bar{\omega}_{l})\vdots
\bar{\mathbf{e}}_{p_{j}\varrho_{j}}(\mathbf{r},\bar{\omega}_{j})\bar{\mathbf{e}}_{q_{k}\tau_{k}}^{*}(\mathbf{r},\bar{\omega}_{k})
\bar{\mathbf{e}}_{m_{l}\sigma_{l}}(\mathbf{r},\bar{\omega}_{l})
A_{p_{j}\varrho_{j}}^{(j)}A_{q_{k}\tau_{k}}^{(k)*}A_{m_{l}\sigma_{l}}^{(l)} \notag\\
&~~~~\left.\times\frac{e^{i\left[(\varrho_{j}\bar{\beta}_{p_{j}}-\tau_{k}\bar{\beta}_{q_{k}}+\sigma_{l}\bar{\beta}_{m_{l}})z
-(\bar{\omega}_{j}-\bar{\omega}_{k}+\bar{\omega}_{l})t\right]}}{\sqrt{\bar{P}_{p_{j}}\bar{P}_{q_{k}}\bar{P}_{m_{l}}}}
\right\vert_{\scriptsize{\begin{array}{lc}
        j\neq k\neq l\neq i \\
        \bar{\omega}_{j}-\bar{\omega}_{k}+\bar{\omega}_{l}=\bar{\omega}_{i}
      \end{array}}}.
\end{align}
\end{widetext}

This expression for the nonlinear polarization accounts for the fact that the nonlinear
susceptibility is invariant to frequency permutations. The first term in Eq.~\eqref{polnltime}
represents SPM effects of the pulse envelopes, the second and third terms describe the XPM between
modes with the same frequency and XPM between pulses propagating at different frequencies,
respectively, whereas the last term describes FWM processes.

If one inserts in Eq.~\eqref{cmetime} the linear and nonlinear polarizations given by
Eq.~\eqref{linp} and Eq.~\eqref{polnltime}, respectively, then discards the fast time-varying
terms, one obtains the following system of coupled equations that governs the dynamics of the mode
envelopes:
\begin{align}\label{cmetimefin}
&\rho_{i}\frac{\displaystyle \partial A_{n_{i}\rho_{i}}^{(i)}}{\displaystyle \partial z} =
i\sum_{q\geq1}\frac{\beta^{(q)i}_{n_{i}\rho_{i}}}{q!}\left(i\frac{\partial}{\partial
t}\right)^{q}A_{n_{i}\rho_{i}}^{(i)} \notag \\
&+i\sum_{q\geq1}\sum\limits_{\scriptsize{(m_{i}\sigma_{i})\neq(n_{i}\rho_{i})}}\frac{\beta^{(q)ii}_{n_{i}\rho_{i},m_{i}\sigma_{i}}}{q!}\left(i\frac{\partial}{\partial t}\right)^{q}A_{m_{i}\sigma_{i}}^{(i)} \notag \\
&+i\frac{\vartheta_{n_{i}\rho_{i}}^{i}(z)}{v_{g,n_{i}}^{i}}A_{n_{i}\rho_{i}}^{(i)}+i\sum\limits_{\scriptsize{(m_{i}\sigma_{i})\neq(n_{i}\rho_{i})}}\frac{\vartheta_{n_{i}\rho_{i},m_{i}\sigma_{i}}^{i}(z)}{\sqrt{v_{g,n_{i}}^{i}v_{g,m_{i}}^{i}}}A_{m_{i}\sigma_{i}}^{(i)} \notag \\
&+\frac{3i\bar{\omega}_{i}}{16\epsilon_{0}a^{2}}\Bigg\{\sum_{m_{i}\sigma_{i}}\Bigg[\frac{\Gamma_{n_{i}\rho_{i},m_{i}\sigma_{i}}^{i}(z)}{v_{g,m_{i}}^{i}\sqrt{v_{g,m_{i}}^{i}v_{g,n_{i}}^{i}}}
\vert
A_{m_{i}\sigma_{i}}^{(i)}{\vert}^{2}A_{m_{i}\sigma_{i}}^{(i)} \notag \\
&+\sum\limits_{\scriptsize{\begin{array}{c}
        (p_{i}\varrho_{i})\neq(m_{i}\sigma_{i}) \\
        p_{i}>m_{i}
      \end{array}}}\frac{2\Gamma_{n_{i}\rho_{i},m_{i}\sigma_{i}p_{i}\varrho_{i}}^{i}(z)}{v_{g,p_{i}}^{i}\sqrt{v_{g,m_{i}}^{i}v_{g,n_{i}}^{i}}}
\vert
A_{p_{i}\varrho_{i}}^{(i)}{\vert}^{2}A_{m_{i}\sigma_{i}}^{(i)} \notag \\
&+\sum\limits_{\scriptsize{\begin{array}{c}
        j=1 \\
        j\neq i
      \end{array}}}^{4}\sum\limits_{\scriptsize{p_{j}\varrho_{j}}}\frac{2\Gamma_{n_{i}\rho_{i},m_{i}\sigma_{i}p_{j}\varrho_{j}}^{ij}(z)}{v_{g,p_{j}}^{j}\sqrt{v_{g,m_{i}}^{i}v_{g,n_{i}}^{i}}}
\vert
A_{p_{j}\varrho_{j}}^{(j)}{\vert}^{2}A_{m_{i}\sigma_{i}}^{(i)}\Bigg] \notag \\
&+\sum\limits_{\scriptsize{\begin{array}{c}
        p_{j}q_{k}m_{l} \\
        \varrho_{j}\tau_{k}\sigma_{l}
      \end{array}}}e^{i\Delta\bar{\beta}_{n_{i}p_{j}q_{k}m_{l}}z}
      \frac{2\Gamma_{n_{i}\rho_{i},p_{j}\varrho_{j}q_{k}\tau_{k}m_{l}\sigma_{l}}^{jkl}(z)}{\sqrt{v_{g,p_{j}}^{j}v_{g,q_{k}}^{k}v_{g,m_{l}}^{l}v_{g,n_{i}}^{i}}}\notag \\
&~~~~~\left.\times
A_{p_{j}\varrho_{j}}^{(j)}A_{q_{k}\tau_{k}}^{(k)*}A_{m_{l}\sigma_{l}}^{(l)}\right\vert_{j\neq
k\neq l\neq i}\Bigg\},~i=1,\ldots,4,
\end{align}
where
$\Delta\bar{\beta}_{n_{i}p_{j}q_{k}m_{l}}=\varrho_{j}\bar{\beta}_{p_{j}}-\tau_{k}\bar{\beta}_{q_{k}}+\sigma_{l}\bar{\beta}_{m_{l}}-\rho_{i}\bar{\beta}_{n_{i}}$
is the wavevector mismatch.

The coefficients $\vartheta_{n_{i}\rho_{i}}^{i}$ and
$\vartheta_{n_{i}\rho_{i},m_{i}\sigma_{i}}^{i}$ represent the wavevector shift of the optical mode
$(n_{i},\rho_{i})$ and the linear coupling constant between modes $(n_{i},\rho_{i})$ and
$(m_{i},\sigma_{i})$, induced by the linear perturbations, respectively,
$\Gamma_{n_{i}\rho_{i},m_{i}\sigma_{i}}^{i}$ and
$\Gamma_{n_{i}\rho_{i},m_{i}\sigma_{i}p_{i}\varrho_{i}}^{i}$ describe SPM and XPM-induced coupling
between modes with the same frequency, $\bar{\omega}_{i}$, respectively,
$\Gamma_{n_{i}\rho_{i},m_{i}\sigma_{i}p_{j}\varrho_{j}}^{ij}$ represents the XPM-induced coupling
between modes with frequencies $\bar{\omega}_{i}$ and $\bar{\omega}_{j}$, and
$\Gamma_{n_{i}\rho_{i},p_{j}\varrho_{j}q_{k}\tau_{k}m_{l}\sigma_{l}}^{jkl}$ is related to the FWM
interaction among the pulses. All these nonlinear coefficients have the meaning of $z$-dependent
effective cubic susceptibilities. The linear and nonlinear coefficients in Eqs.~\eqref{cmetimefin}
are given by the following relations:
\begin{subequations}\label{couplcoefflinnl}
\begin{align}
\label{couplcoeffLinS}
&\vartheta_{n_{i}\rho_{i}}^{i}(z) = \frac{\bar{\omega}_{i}a}{4\bar{W}_{n_{i}}^{i}}\int_{S}[\delta \epsilon_{\mathrm{fc}}(\mathbf{r})+\delta \epsilon_{\mathrm{loss}}(\mathbf{r})]\vert\mathbf{e}_{n_{i}\rho_{i}}(\bar{\omega}_{i})\vert^{2}dS, \\
\label{couplcoeffLinX} &\vartheta_{n_{i}\rho_{i},m_{i}\sigma_{i}}^{i}(z) =
\frac{\bar{\omega}_{i}e^{i(\sigma_{i}\bar{\beta}_{m_{i}}-\rho_{i}\bar{\beta}_{n_{i}})z}}{4\sqrt{\bar{W}_{n_{i}}^{i}\bar{W}_{m_{i}}^{i}}} \notag \\
&~~\times\int_{S}[\delta \epsilon_{\mathrm{fc}}(\mathbf{r})+\delta
\epsilon_{\mathrm{loss}}(\mathbf{r})]\mathbf{e}_{n_{i}\rho_{i}}^{*}(\bar{\omega}_{i})\cdot\mathbf{e}_{m_{i}\sigma_{i}}(\bar{\omega}_{i})dS, \\
\label{GammaSPM} &\Gamma_{n_{i}\rho_{i},m_{i}\sigma_{i}}^{i}(z) =
\frac{\epsilon_{0}^{2}a^{4}e^{i(\sigma_{i}\bar{\beta}_{m_{i}}-\rho_{i}\bar{\beta}_{n_{i}})z}}{\bar{W}_{m_{i}}^{i}\sqrt{\bar{W}_{m_{i}}^{i}\bar{W}_{n_{i}}^{i}}}\int_{S}\mathbf{e}_{n_{i}\rho_{i}}^{*}(\bar{\omega}_{i}) \notag \\
&~~\cdot\hat{\chi}^{(3)}(\bar{\omega}_{i},-\bar{\omega}_{i},\bar{\omega}_{i})\vdots\mathbf{e}_{m_{i}\sigma_{i}}(\bar{\omega}_{i})\mathbf{e}_{m_{i}\sigma_{i}}^{*}(\bar{\omega}_{i})\mathbf{e}_{m_{i}\sigma_{i}}(\bar{\omega}_{i})dS, \\
\label{GammaXPM} &\Gamma_{n_{i}\rho_{i},m_{i}\sigma_{i}p_{i}\varrho_{i}}^{i}(z) =
\frac{\epsilon_{0}^{2}a^{4}e^{i(\sigma_{i}\bar{\beta}_{m_{i}}-\rho_{i}\bar{\beta}_{n_{i}})z}}{\bar{W}_{p_{i}}^{i}\sqrt{\bar{W}_{m_{i}}^{i}\bar{W}_{n_{i}}^{i}}}\int_{S}\mathbf{e}_{n_{i}\rho_{i}}^{*}(\bar{\omega}_{i}) \notag \\
&~~\cdot\hat{\chi}^{(3)}(\bar{\omega}_{i},-\bar{\omega}_{i},\bar{\omega}_{i})\vdots\mathbf{e}_{p_{i}\varrho_{i}}(\bar{\omega}_{i})\mathbf{e}_{p_{i}\varrho_{i}}^{*}(\bar{\omega}_{i})\mathbf{e}_{m_{i}\sigma_{i}}(\bar{\omega}_{i})dS, \\
\label{GammaXPMdiffFreq} &\Gamma_{n_{i}\rho_{i},m_{i}\sigma_{i}p_{j}\varrho_{j}}^{ij}(z) =
\frac{\epsilon_{0}^{2}a^{4}e^{i(\sigma_{i}\bar{\beta}_{m_{i}}-\rho_{i}\bar{\beta}_{n_{i}})z}}{\bar{W}_{p_{j}}^{j}\sqrt{\bar{W}_{m_{i}}^{i}\bar{W}_{n_{i}}^{i}}}\int_{S}\mathbf{e}_{n_{i}\rho_{i}}^{*}(\bar{\omega}_{i}) \notag \\
&~~\cdot\hat{\chi}^{(3)}(\bar{\omega}_{j},-\bar{\omega}_{j},\bar{\omega}_{i})\vdots\mathbf{e}_{p_{j}\varrho_{j}}(\bar{\omega}_{j})\mathbf{e}_{p_{j}\varrho_{j}}^{*}(\bar{\omega}_{j})\mathbf{e}_{m_{i}\sigma_{i}}(\bar{\omega}_{i})dS, \\
\label{GammaFWM} &\Gamma_{n_{i}\rho_{i},p_{j}\varrho_{j}q_{k}\tau_{k}m_{l}\sigma_{l}}^{jkl}(z) =
\frac{\epsilon_{0}^{2}a^{4}}{\sqrt{\bar{W}_{p_{j}}^{j}\bar{W}_{q_{k}}^{k}\bar{W}_{m_{l}}^{l}\bar{W}_{n_{i}}^{i}}}\int_{S}\mathbf{e}_{n_{i}\rho_{i}}^{*}(\bar{\omega}_{i}) \notag \\
&~~\cdot\hat{\chi}^{(3)}(\bar{\omega}_{j},-\bar{\omega}_{k},\bar{\omega}_{l})\vdots\mathbf{e}_{p_{j}\varrho_{j}}(\bar{\omega}_{j})\mathbf{e}_{q_{k}\tau_{k}}^{*}(\bar{\omega}_{k})\mathbf{e}_{m_{l}\sigma_{l}}(\bar{\omega}_{l})dS.
\end{align}
\end{subequations}

While Eqs.~\eqref{cmetimefin} seem complicated, in cases of practical interest they can be
considerably simplified. To be more specific, these equations describe a multitude of optical
effects pertaining to both linear and nonlinear gratings, including linear coupling between modes
with the same frequency, nonlinear coupling between modes with the same frequency, due to SPM and
XPM effects, XPM-induced coupling between modes with different frequency, and FWM interactions. In
most experimental set-ups, however, not all these linear and nonlinear effects occur
simultaneously as in a generic case not all of them lead to efficient pulse interactions.

These ideas becomes clear if one inspects the exponential factors in
Eqs.~\eqref{couplcoeffLinX}-\eqref{GammaXPMdiffFreq}. Thus, they vary over a characteristic length
comparable to the lattice constant of the PhC, namely much more rapidly as compared to the spatial
variation rate of the pulse envelopes. As a result, except for the mode $(n_{i},\rho_{i})$, these
linear and nonlinear coefficients cancel. There are, however, particular cases when some of these
interactions are phase-matched and consequently are resonantly enhanced. To be more specific, the
integrals in Eqs.~\eqref{couplcoeffLinX}-\eqref{GammaXPMdiffFreq} are periodic functions of $z$,
with period $a$, so that it is possible that a Fourier component of these integrals phase-matches
a specific linear or nonlinear interaction between modes (e.g., the linear coupling between two
modes with the same frequency and SPM- or XPM-induced nonlinear coupling between modes). In this
study, we do not consider such accidental phase-matching of mode interactions. With this in mind,
we discard all terms in Eqs.~\eqref{cmetimefin} that average to zero to obtain the final form of
the coupled-mode equations for the pulse envelopes:
\begin{align}\label{cmetimefinav}
i&\bigg[\rho_{i}\frac{\displaystyle \partial A_{n_{i}\rho_{i}}^{(i)}}{\displaystyle \partial z}
+\frac{\delta^{i}_{n_{i}}(z)}{v_{g,n_{i}}^{i}}\frac{\partial A_{n_{i}\rho_{i}}^{(i)}}{\partial
t}\bigg]-\frac{\delta^{i}_{n_{i}}(z)\bar{\beta}_{2,n_{i}}}{2}\frac{\partial^{2}
A_{n_{i}\rho_{i}}^{(i)}}{\partial
t^{2}}\notag \\
&+\frac{\bar{\omega}_{i}\delta n_{\mathrm{fc}}\bar{\kappa}_{n_{i}}^{i}(z)}{n v_{g,n_{i}}^{i}}A_{n_{i}\rho_{i}}^{(i)}+\frac{ic\bar{\kappa}_{n_{i}}^{i}(z)}{2n v_{g,n_{i}}^{i}}(\alpha_{\mathrm{fc}}+\alpha_{\mathrm{in}})A_{n_{i}\rho_{i}}^{(i)} \notag \\
&+\gamma_{n_{i}\rho_{i}}^{i}(z)\vert A_{n_{i}\rho_{i}}^{(i)}{\vert}^{2}A_{n_{i}\rho_{i}}^{(i)}
+\sum\limits_{\scriptsize{\begin{array}{c}
        (p_{i}\varrho_{i})\neq(n_{i}\rho_{i}) \\
        p_{i}>n_{i}
      \end{array}}}2\gamma_{n_{i}\rho_{i},p_{i}\varrho_{i}}^{i}(z) \notag \\
&\times\vert
A_{p_{i}\varrho_{i}}^{(i)}{\vert}^{2}A_{n_{i}\rho_{i}}^{(i)}+\sum\limits_{\scriptsize{\begin{array}{c}
        j=1 \\
        j\neq i
      \end{array}}}^{4}\sum\limits_{\scriptsize{p_{j}\varrho_{j}}}2\gamma_{n_{i}\rho_{i},p_{j}\varrho_{j}}^{ij}(z)
\vert A_{p_{j}\varrho_{j}}^{(j)}{\vert}^{2}A_{n_{i}\rho_{i}}^{(i)} \notag \\
&+\sum\limits_{\scriptsize{\begin{array}{c}
        p_{j}q_{k}m_{l} \\
        \varrho_{j}\tau_{k}\sigma_{l}
      \end{array}}}2e^{i\Delta\bar{\beta}_{n_{i}p_{j}q_{k}m_{l}}z}\gamma_{n_{i}\rho_{i},p_{j}\varrho_{j}q_{k}\tau_{k}m_{l}\sigma_{l}}^{jkl}(z) \notag \\
&~~~~~\left.\times
A_{p_{j}\varrho_{j}}^{(j)}A_{q_{k}\tau_{k}}^{(k)*}A_{m_{l}\sigma_{l}}^{(l)}\right\vert_{j\neq
k\neq l\neq i}=0,~i=1,\ldots,4,
\end{align}
where the new parameters introduced in this equation are defined as:
\begin{subequations}\label{couplcoefflinnlav}
\begin{align}
\label{couplcoeffLinS}
&\bar{\kappa}_{n_{i}}^{i}(z) = \frac{\epsilon_{0}an^{2}}{2\bar{W}_{n_{i}}^{i}}\int_{S_{\mathrm{nl}}}\vert\mathbf{e}_{n_{i}\rho_{i}}(\bar{\omega}_{i})\vert^{2}dS, \\
\label{GammaSPMav} &\gamma_{n_{i}\rho_{i}}^{i}(z) =
\frac{3\bar{\omega}_{i}\epsilon_{0}a^{2}}{16v_{g,n_{i}}^{i^{\scriptstyle 2}}}\frac{1}{\bar{W}_{n_{i}}^{i^{\scriptstyle 2}}}\int_{S_{\mathrm{nl}}}\mathbf{e}_{n_{i}\rho_{i}}^{*}(\bar{\omega}_{i}) \notag \\
&~~\cdot\hat{\chi}^{(3)}(\bar{\omega}_{i},-\bar{\omega}_{i},\bar{\omega}_{i})\vdots\mathbf{e}_{n_{i}\rho_{i}}(\bar{\omega}_{i})\mathbf{e}_{n_{i}\rho_{i}}^{*}(\bar{\omega}_{i})\mathbf{e}_{n_{i}\rho_{i}}(\bar{\omega}_{i})dS, \\
\label{GammaXPMav} &\gamma_{n_{i}\rho_{i},p_{i}\varrho_{i}}^{i}(z) =
\frac{3\bar{\omega}_{i}\epsilon_{0}a^{2}}{16v_{g,n_{i}}^{i}v_{g,p_{i}}^{i}}\frac{1}{\bar{W}_{n_{i}}^{i}\bar{W}_{p_{i}}^{i}}\int_{S_{\mathrm{nl}}}\mathbf{e}_{n_{i}\rho_{i}}^{*}(\bar{\omega}_{i}) \notag \\
&~~\cdot\hat{\chi}^{(3)}(\bar{\omega}_{i},-\bar{\omega}_{i},\bar{\omega}_{i})\vdots\mathbf{e}_{p_{i}\varrho_{i}}(\bar{\omega}_{i})\mathbf{e}_{p_{i}\varrho_{i}}^{*}(\bar{\omega}_{i})\mathbf{e}_{n_{i}\rho_{i}}(\bar{\omega}_{i})dS, \\
\label{GammaXPMdiffFreqav} &\gamma_{n_{i}\rho_{i},p_{j}\varrho_{j}}^{ij}(z) =
\frac{3\bar{\omega}_{i}\epsilon_{0}a^{2}}{16v_{g,n_{i}}^{i}v_{g,p_{j}}^{j}}\frac{1}{\bar{W}_{n_{i}}^{i}\bar{W}_{p_{j}}^{j}}\int_{S_{\mathrm{nl}}}\mathbf{e}_{n_{i}\rho_{i}}^{*}(\bar{\omega}_{i}) \notag \\
&~~\cdot\hat{\chi}^{(3)}(\bar{\omega}_{j},-\bar{\omega}_{j},\bar{\omega}_{i})\vdots\mathbf{e}_{p_{j}\varrho_{j}}(\bar{\omega}_{j})\mathbf{e}_{p_{j}\varrho_{j}}^{*}(\bar{\omega}_{j})\mathbf{e}_{n_{i}\rho_{i}}(\bar{\omega}_{i})dS, \\
\label{GammaFWMav} &\gamma_{n_{i}\rho_{i},p_{j}\varrho_{j}q_{k}\tau_{k}m_{l}\sigma_{l}}^{jkl}(z) =
\frac{3\bar{\omega}_{i}\epsilon_{0}a^{2}}{16{(v_{g,p_{j}}^{j}v_{g,q_{k}}^{k}v_{g,m_{l}}^{l}v_{g,n_{i}}^{i})}^{\frac{1}{2}}} \notag \\
&~~\frac{1}{{(\bar{W}_{p_{j}}^{j}\bar{W}_{q_{k}}^{k}\bar{W}_{m_{l}}^{l}\bar{W}_{n_{i}}^{i})}^{\frac{1}{2}}}\int_{S_{\mathrm{nl}}}\mathbf{e}_{n_{i}\rho_{i}}^{*}(\bar{\omega}_{i})\cdot
\hat{\chi}^{(3)}(\bar{\omega}_{j},-\bar{\omega}_{k},\bar{\omega}_{l}) \notag \\
&~~\vdots\mathbf{e}_{p_{j}\varrho_{j}}(\bar{\omega}_{j})\mathbf{e}_{q_{k}\tau_{k}}^{*}(\bar{\omega}_{k})\mathbf{e}_{m_{l}\sigma_{l}}(\bar{\omega}_{l})dS.
\end{align}
\end{subequations}

In these equations, $S_{\mathrm{nl}}(z)$ is the transverse surface of the region filled with
nonlinear material. Note that the exponential factor in the term describing the FWM does not
average to zero because the FWM interaction is assumed to be nearly phase-matched and therefore
the exponential factor varies over a characteristic length that is much larger than the lattice
constant, $a$. Importantly, the linear and nonlinear effects in Eq.~\eqref{cmetimefinav} appear as
being inverse proportional to the $v_{g}$ and $v_{g}^{2}$, respectively. In other words, one does
not need to rely on any phenomenological considerations to describe slow-light effects, as they
are naturally captured by our model.

\subsection{Carriers dynamics} The last step in our derivation of the theoretical model describing
FWM in Si-PhCWGs is to determine the influence of photogenerated FCs on pulse dynamics. To this
end, we first find the rate at which electron-hole pairs are generated optically, \textit{via}
degenerate and nondegenerate TPA, and as a result of FWM. More specifically, we first multiply
Eqs.~\eqref{cmetimefin}, after all linear terms have been discarded, by
$A_{n_{i}\rho_{i}}^{(i)*}$, then multiply the complex conjugate of Eqs.~\eqref{cmetimefin} by
$A_{n_{i}\rho_{i}}^{(i)}$, and sum the results over all carrier frequencies and modes. The outcome
of these simple manipulations can be cast as:
\begin{align}\label{PowTPAXAMFWMtime}
&\frac{\displaystyle \partial}{\displaystyle \partial
z}\sum_{i=1}^{4}\sum_{n_{i}\rho_{i}}\rho_{i}\vert
A_{n_{i}\rho_{i}}^{(i)}\vert^{2}=- \frac{3}{8\epsilon_{0}a^{2}}\sum_{i=1}^{4}\sum_{n_{i}\rho_{i}}\bar{\omega}_{i}\notag \\
&\times\mathfrak{Im}\Bigg\{\sum_{m_{i}\sigma_{i}}\Bigg[\frac{\Gamma_{n_{i}\rho_{i},m_{i}\sigma_{i}}^{i}(z)}{v_{g,m_{i}}^{i}\sqrt{v_{g,m_{i}}^{i}v_{g,n_{i}}^{i}}}
\vert
A_{m_{i}\sigma_{i}}^{(i)}{\vert}^{2}A_{m_{i}\sigma_{i}}^{(i)}A_{n_{i}\rho_{i}}^{(i)*} \notag \\
&+\sum\limits_{\scriptsize{\begin{array}{c}
        (p_{i}\varrho_{i})\neq(m_{i}\sigma_{i}) \\
        p_{i}>m_{i}
      \end{array}}}\frac{2\Gamma_{n_{i}\rho_{i},m_{i}\sigma_{i}p_{i}\varrho_{i}}^{i}(z)}{v_{g,p_{i}}^{i}\sqrt{v_{g,m_{i}}^{i}v_{g,n_{i}}^{i}}}
\vert
A_{p_{i}\varrho_{i}}^{(i)}{\vert}^{2}A_{m_{i}\sigma_{i}}^{(i)}A_{n_{i}\rho_{i}}^{(i)*} \notag \\
&+\sum\limits_{\scriptsize{\begin{array}{c}
        j=1 \\
        j\neq i
      \end{array}}}^{4}\sum\limits_{\scriptsize{p_{j}\varrho_{j}}}\frac{2\Gamma_{n_{i}\rho_{i},m_{i}\sigma_{i}p_{j}\varrho_{j}}^{ij}(z)}{v_{g,p_{j}}^{j}\sqrt{v_{g,m_{i}}^{i}v_{g,n_{i}}^{i}}}
\vert
A_{p_{j}\varrho_{j}}^{(j)}{\vert}^{2}A_{m_{i}\sigma_{i}}^{(i)}A_{n_{i}\rho_{i}}^{(i)*}\Bigg] \notag \\
&+\sum\limits_{\scriptsize{\begin{array}{c}
        p_{j}q_{k}m_{l} \\
        \varrho_{j}\tau_{k}\sigma_{l}
      \end{array}}}e^{i\Delta\bar{\beta}_{n_{i}p_{j}q_{k}m_{l}}z}
      \frac{2\Gamma_{n_{i}\rho_{i},p_{j}\varrho_{j}q_{k}\tau_{k}m_{l}\sigma_{l}}^{jkl}(z)}{\sqrt{v_{g,p_{j}}^{j}v_{g,q_{k}}^{k}v_{g,m_{l}}^{l}v_{g,n_{i}}^{i}}}\notag \\
&~~~~~\left.\times
A_{p_{j}\varrho_{j}}^{(j)}A_{q_{k}\tau_{k}}^{(k)*}A_{m_{l}\sigma_{l}}^{(l)}A_{n_{i}\rho_{i}}^{(i)*}\right\vert_{j\neq
k\neq l\neq i}\Bigg\}.
\end{align}

The sum in the l.h.s. of this equation represents the rate at which optical power is transferred
to FCs. This power is absorbed by carriers generated in the silicon slab, in the infinitesimal
volume $dV(z)=A_{\mathrm{nl}}(z)dz$, where $A_{\mathrm{nl}}(z)$ is an effective area. This area is
defined in terms of the Poynting vector of the field propagating inside the silicon slab,
\begin{align}\label{effArea}
A_{\mathrm{nl}}(z) = \frac{\big[\int_{S_{\mathrm{nl}}}\vert\langle
\mathbf{E}(\mathbf{r},t)\times\mathbf{H}(\mathbf{r},t)\rangle_{t}\vert
dS\big]^{2}}{\int_{S_{\mathrm{nl}}}\vert\langle
\mathbf{E}(\mathbf{r},t)\times\mathbf{H}(\mathbf{r},t)\rangle_{t}\vert^{2} dS}.
\end{align}
In this equation, $\langle f\rangle_{t}$ means the time average of $f$. Using
Eq.~\eqref{fieldtimesimpl}, and taking into account the fact that $A_{n_{i}\rho_{i}}^{(i)}$ varies
in time much slower than $e^{-i\bar{\omega}_{i} t}$, one can express Eq.~\eqref{effArea} in the
following form:
\begin{align}\label{effAreaSimpl}
A_{\mathrm{nl}}(z) = \frac{\displaystyle
\bigg[\int_{S_{\mathrm{nl}}}\bigg\lvert\sum_{i=1}^{4}\sum_{n_{i}\rho_{i}}\frac{\vert
A_{n_{i}\rho_{i}}^{(i)}\vert^{2}}{\bar{P}_{n_{i}}}
\mathfrak{Re}(\mathbf{e}_{n_{i}\rho_{i}}\times\mathbf{h}_{n_{i}\rho_{i}}^{*})\bigg\rvert
dS\bigg]^{2}}{\displaystyle
\int_{S_{\mathrm{nl}}}\bigg\lvert\sum_{i=1}^{4}\sum_{n_{i}\rho_{i}}\frac{\vert
A_{n_{i}\rho_{i}}^{(i)}\vert^{2}}{\bar{P}_{n_{i}}}
\mathfrak{Re}(\mathbf{e}_{n_{i}\rho_{i}}\times\mathbf{h}_{n_{i}\rho_{i}}^{*})\bigg\rvert^{2} dS}.
\end{align}

In spite of the fact that it might seem difficult to use this formula to calculate the effective
area, we will show in the next section that in cases of practical interest it can be simplified
considerably. We also stress that Eq.~\eqref{effAreaSimpl} gives the effective transverse area of
the region in which FCs are generated, so that it should not be confused with the modal effective
area. In fact, since in the FWM process there are several co-propagating beams, a single effective
modal area is not well defined.

The energy transferred to FCs when an electron-hole pair is generated \textit{via} absorption of
two photons with frequencies $\bar{\omega}_{i}$ and $\bar{\omega}_{j}$ is equal to
$\hbar(\bar{\omega}_{i}+\bar{\omega}_{j})$. Using this result and neglecting again all terms in
Eq.~\eqref{PowTPAXAMFWMtime} that average to zero, it can be easily shown that the carriers
dynamics are governed by the following rate equation:
\begin{align}\label{FCdyn}
&\frac{\partial N}{\partial t} = -\frac{N}{\tau_{c}}+\frac{1}{\hbar
A_{\mathrm{nl}}(z)}\sum_{i=1}^{4}\sum_{n_{i}\rho_{i}}\bigg\{\frac{\gamma_{n_{i}\rho_{i}}^{\prime\prime
i}(z)}{\bar{\omega}_{i}}\vert
A_{n_{i}\rho_{i}}^{(i)}{\vert}^{4} \notag \\
&~~~+\sum\limits_{\scriptsize{\begin{array}{c}
        (p_{i}\varrho_{i})\neq(n_{i}\rho_{i}) \\
        p_{i}>n_{i}
      \end{array}}}\frac{2\gamma_{n_{i}\rho_{i},p_{i}\varrho_{i}}^{\prime\prime i}(z)}{\bar{\omega}_{i}}\vert A_{p_{i}\varrho_{i}}^{(i)}{\vert}^{2}\vert
A_{n_{i}\rho_{i}}^{(i)}{\vert}^{2} \notag \\
&~~~+\sum\limits_{\scriptsize{\begin{array}{c}
        j=1 \\
        j\neq i
      \end{array}}}^{4}\sum\limits_{\scriptsize{p_{j}\varrho_{j}}}\frac{4\gamma_{n_{i}\rho_{i},p_{j}\varrho_{j}}^{\prime\prime ij}(z)}{\bar{\omega}_{i}+\bar{\omega}_{j}}
\vert A_{p_{j}\varrho_{j}}^{(j)}{\vert}^{2}\vert A_{n_{i}\rho_{i}}^{(i)}{\vert}^{2} \notag \\
&~~~+\sum\limits_{\scriptsize{\begin{array}{c}
        p_{j}q_{k}m_{l} \\
        \varrho_{j}\tau_{k}\sigma_{l}
      \end{array}}}\mathfrak{Im}\bigg[e^{i\Delta\bar{\beta}_{n_{i}p_{j}q_{k}m_{l}}z}\frac{4\gamma_{n_{i}\rho_{i},p_{j}\varrho_{j}q_{k}\tau_{k}m_{l}\sigma_{l}}^{jkl}(z)}{\bar{\omega}_{i}+\bar{\omega}_{k}} \notag \\
&~~~~~~~\left.\times
A_{p_{j}\varrho_{j}}^{(j)}A_{q_{k}\tau_{k}}^{(k)*}A_{m_{l}\sigma_{l}}^{(l)}A_{n_{i}\rho_{i}}^{(i)*}\right\vert_{j\neq
k\neq l\neq i}\bigg]\bigg\},
\end{align}
where $\tau_{c}\approx\SI{500}{\pico\second}$ \cite{mch09oe} is the FC recombination time in
Si-PhCWGs and $\zeta^{\prime}$ ($\zeta^{\prime\prime}$) means the real (imaginary) part of the
complex number, $\zeta$.

\section{Degenerate four-wave mixing}\label{sDFWM}
The system of coupled nonlinear partial differential equations, Eqs.~\eqref{cmetimefinav} and
Eq.~\eqref{FCdyn}, fully describes the FWM of optical pulses and FCs dynamics and represents the
main result derived in this study. In practical experimental set-ups, however, the most used pulse
configuration is that of degenerate FWM. In this particular case, the optical frequencies of the
two pump pulses are the same, $\bar{\omega}_{1}=\bar{\omega}_{2}\equiv\omega_{p}$, whereas the two
generated pulses, the signal and the idler, have frequencies $\bar{\omega}_{3}\equiv\omega_{s}$
and $\bar{\omega}_{4}\equiv\omega_{i}$, respectively. Moreover, we assume that all modes are
forward-propagating modes and that at each carrier frequency there is only one guided mode in
which the optical pulses that enter in the FWM process can propagate -- others, should they exist,
would not be phase-matched -- so that we set $N_{i}=1$, $i=1,\ldots,4$. Under these circumstances,
Eqs.~\eqref{cmetimefinav} and Eq.~\eqref{FCdyn} can be simplified to:
\begin{subequations}\label{FWMpsi}
\begin{align}\label{FWMp}
i&\bigg[\frac{\displaystyle \partial A_{p}}{\displaystyle \partial z}
+\frac{\delta_{p}(z)}{v_{g,p}}\frac{\partial A_{p}}{\partial
t}\bigg]-\frac{\delta_{p}(z)\bar{\beta}_{2,p}}{2}\frac{\partial^{2} A_{p}}{\partial
t^{2}}\notag \\
&+\frac{\omega_{p}\delta n_{\mathrm{fc}}\bar{\kappa}_{p}(z)}{n v_{g,p}}A_{p}+\frac{ic\bar{\kappa}_{p}(z)}{2n v_{g,p}}(\alpha_{\mathrm{fc}}+\alpha_{\mathrm{in}})A_{p} \notag \\
&+\left[\gamma_{p}(z)\vert A_{p}{\vert}^{2}+2\gamma_{ps}(z)\vert
A_{s}{\vert}^{2}+2\gamma_{pi}(z)\vert A_{i}{\vert}^{2}\right]A_{p} \notag \\
&+2e^{i\Delta\bar{\beta}z}\gamma_{psi}(z)A_{s}A_{i}A_{p}^{*}=0, \\\label{FWMs}
i&\bigg[\frac{\displaystyle
\partial A_{s}}{\displaystyle \partial z} +\frac{\delta_{s}(z)}{v_{g,s}}\frac{\partial
A_{s}}{\partial t}\bigg]-\frac{\delta_{s}(z)\bar{\beta}_{2,s}}{2}\frac{\partial^{2}
A_{s}}{\partial t^{2}}\notag \\
&+\frac{\omega_{s}\delta n_{\mathrm{fc}}\bar{\kappa}_{s}(z)}{n v_{g,s}}A_{s}+\frac{ic\bar{\kappa}_{s}(z)}{2n v_{g,s}}(\alpha_{\mathrm{fc}}+\alpha_{\mathrm{in}})A_{s} \notag \\
&+\left[\gamma_{s}(z)\vert A_{s}{\vert}^{2}+2\gamma_{sp}(z)\vert
A_{p}{\vert}^{2}+2\gamma_{si}(z)\vert A_{i}{\vert}^{2}\right]A_{s} \notag \\
&+e^{-i\Delta\bar{\beta}z}\gamma_{spi}(z)A_{p}^{2}A_{i}^{*}=0, \\\label{FWMi}
i&\bigg[\frac{\displaystyle
\partial A_{i}}{\displaystyle \partial z} +\frac{\delta_{i}(z)}{v_{g,i}}\frac{\partial
A_{i}}{\partial t}\bigg]-\frac{\delta_{i}(z)\bar{\beta}_{2,i}}{2}\frac{\partial^{2}
A_{i}}{\partial t^{2}}\notag \\
&+\frac{\omega_{i}\delta n_{\mathrm{fc}}\bar{\kappa}_{i}(z)}{n v_{g,i}}A_{i}+\frac{ic\bar{\kappa}_{i}(z)}{2n v_{g,i}}(\alpha_{\mathrm{fc}}+\alpha_{\mathrm{in}})A_{i} \notag \\
&+\left[\gamma_{i}(z)\vert A_{i}{\vert}^{2}+2\gamma_{ip}(z)\vert
A_{p}{\vert}^{2}+2\gamma_{is}(z)\vert A_{s}{\vert}^{2}\right]A_{i} \notag \\
&+e^{-i\Delta\bar{\beta}z}\gamma_{ips}(z)A_{p}^{2}A_{s}^{*}=0,
\end{align}
\end{subequations}
\begin{align}\label{FWMFCdyn}
&\frac{\partial N}{\partial t} = -\frac{N}{\tau_{c}}+\frac{1}{\hbar
A_{\mathrm{nl}}(z)}\bigg\{\sum_{\mu=p,s,i}\bigg[\frac{\gamma_{\mu}^{\prime\prime}(z)}{\omega_{\mu}}\vert
A_{\mu}{\vert}^{4} \notag \\
&~+\sum\limits_{\scriptsize{\begin{array}{c}
        \nu=p,s,i \\
        \nu\neq \mu
      \end{array}}}\frac{4\gamma_{\mu\nu}^{\prime\prime}(z)}{\omega_{\mu}+\omega_{\nu}}
\vert A_{\mu}{\vert}^{2}\vert
A_{\nu}{\vert}^{2}\bigg]+\frac{1}{\omega_{p}}\mathfrak{Im}\big[2\gamma_{psi}(z) \notag \\
&~\times {A_{p}^{*}}^{2}A_{s}A_{i}e^{i\Delta\bar{\beta}z}+[\gamma_{spi}(z)+\gamma_{ips}(z)]
A_{p}^{2}A_{s}^{*}A_{i}^{*}e^{-i\Delta\bar{\beta}z}\big]\big\},
\end{align}
where $\Delta\bar{\beta}=\beta_s+\beta_i-2\beta_p$. The coefficients of the linear and nonlinear
terms in Eqs.~\eqref{FWMpsi} and Eq.~\eqref{FWMFCdyn} are:
\begin{figure}[b]
\centerline{\includegraphics[width=8cm]{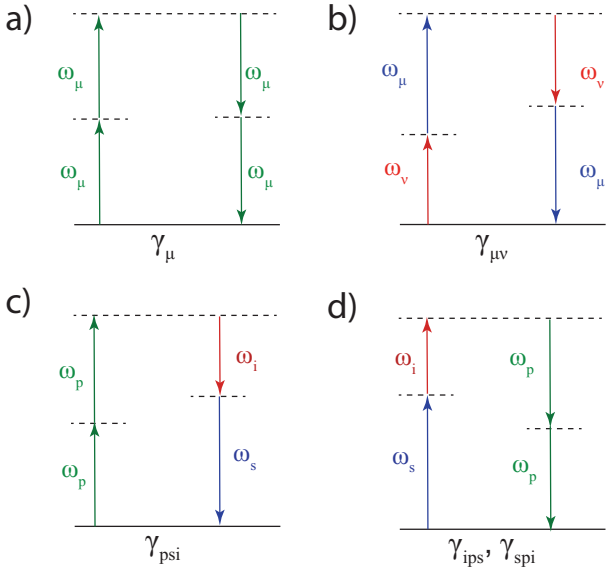}} \caption{Energy diagrams representing the
nonlinear optical processes included in Eqs.~\eqref{FWMpsi}. (a) SPM and degenerate TPA
corresponding to $\gamma_{\mu}^{\prime}$ and $\gamma_{\mu}^{\prime\prime}$, respectively. (b) XPM
and XAM corresponding to $\gamma_{\mu\nu}^{\prime}$ and $\gamma_{\mu\nu}^{\prime\prime}$,
respectively. Two possible ways of energy transfer that can occur during a degenerate FWM process:
(c) two pump photons generate a signal and an idler photon, a process described by $\gamma_{psi}$;
(d) the reverse process, described by $\gamma_{ips}$ and $\gamma_{spi}$, in which a signal and an
idler photon generate two pump photons.} \label{fig:endiagr}
\end{figure}
\begin{subequations}\label{FWMcouplcoefflinnlav}
\begin{align}
\label{FWMcouplcoeffLinS}
&\bar{\kappa}_{\mu}(z) = \frac{\epsilon_{0}an^{2}}{2\bar{W}_{\mu}}\int_{S_{\mathrm{nl}}}\vert\mathbf{e}_{\mu}(\omega_{\mu})\vert^{2}dS, \\
\label{FWMGammaSPMav} &\gamma_{\mu}(z) =
\frac{3\omega_{\mu}\epsilon_{0}a^{2}}{16v_{g,\mu}^{2}}\frac{1}{\bar{W}_{\mu}^{2}}\int_{S_{\mathrm{nl}}}\mathbf{e}_{\mu}^{*}(\omega_{\mu}) \notag \\
&~~\cdot\hat{\chi}^{(3)}(\omega_{\mu},-\omega_{\mu},\omega_{\mu})\vdots\mathbf{e}_{\mu}(\omega_{\mu})\mathbf{e}_{\mu}^{*}(\omega_{\mu})\mathbf{e}_{\mu}(\omega_{\mu})dS, \\
\label{FWMGammaXPMdiffFreqav} &\gamma_{\mu\nu}(z) =
\frac{3\omega_{\mu}\epsilon_{0}a^{2}}{16v_{g,\mu}v_{g,\nu}}\frac{1}{\bar{W}_{\mu}\bar{W}_{\nu}}\int_{S_{\mathrm{nl}}}\mathbf{e}_{\mu}^{*}(\omega_{\mu}) \notag \\
&~~\cdot\hat{\chi}^{(3)}(\omega_{\nu},-\omega_{\nu},\omega_{\mu})\vdots\mathbf{e}_{\nu}(\omega_{\nu})\mathbf{e}_{\nu}^{*}(\omega_{\nu})\mathbf{e}_{\mu}(\omega_{\mu})dS, \\
\label{FWMpsiGammaFWMav} &\gamma_{psi}(z) =
\frac{3\omega_{p}\epsilon_{0}a^{2}}{16v_{g,p}{(v_{g,s}v_{g,i})}^{\frac{1}{2}}}\frac{1}{\bar{W}_{p}{(\bar{W}_{s}\bar{W}_{i})}^{\frac{1}{2}}} \notag \\
&~~\times\int_{S_{\mathrm{nl}}}\mathbf{e}_{p}^{*}(\omega_{p})\cdot
\hat{\chi}^{(3)}(\omega_{s},-\omega_{p},\omega_{i})\vdots\mathbf{e}_{s}(\omega_{s})\mathbf{e}_{p}^{*}(\omega_{p})\mathbf{e}_{i}(\omega_{i})dS, \\
\label{FWMspiGammaFWMav} &\gamma_{spi}(z) =
\frac{3\omega_{s}\epsilon_{0}a^{2}}{16v_{g,p}{(v_{g,s}v_{g,i})}^{\frac{1}{2}}}\frac{1}{\bar{W}_{p}{(\bar{W}_{s}\bar{W}_{i})}^{\frac{1}{2}}} \notag \\
&~~\times\int_{S_{\mathrm{nl}}}\mathbf{e}_{s}^{*}(\omega_{s})\cdot
\hat{\chi}^{(3)}(\omega_{p},-\omega_{i},\omega_{p})\vdots\mathbf{e}_{p}(\omega_{p})\mathbf{e}_{i}^{*}(\omega_{i})\mathbf{e}_{p}(\omega_{p})dS, \\
\label{FWMipsGammaFWMav} &\gamma_{ips}(z) =
\frac{3\omega_{i}\epsilon_{0}a^{2}}{16v_{g,p}{(v_{g,s}v_{g,i})}^{\frac{1}{2}}}\frac{1}{\bar{W}_{p}{(\bar{W}_{s}\bar{W}_{i})}^{\frac{1}{2}}} \notag \\
&~~\times\int_{S_{\mathrm{nl}}}\mathbf{e}_{i}^{*}(\omega_{i})\cdot
\hat{\chi}^{(3)}(\omega_{p},-\omega_{s},\omega_{p})\vdots\mathbf{e}_{p}(\omega_{p})\mathbf{e}_{s}^{*}(\omega_{s})\mathbf{e}_{p}(\omega_{p})dS,
\end{align}
\end{subequations}
where $\mu$ and $\nu\neq\mu$ take one of the values $p$, $s$, and $i$ and the frequency degeneracy
at the pump frequency has been taken into account. Note that, as expected, when the nonlinear
coefficients $\gamma$'s are real quantities, namely when nonlinear optical absorption effects can
be neglected, the optical pumping term in Eq.~\eqref{FWMFCdyn} vanishes. Moreover, since in
experiments usually $P_{p}\gg P_{s},P_{i}$, the effective area given by Eq.~\eqref{effAreaSimpl}
can be reduced to the following simplified form:
\begin{equation}\label{effAreaSimplSW}
A_{\mathrm{nl}}(z) = \frac{\displaystyle \left(\int_{S_{\mathrm{nl}}}\big\vert
\mathfrak{Re}\left[\mathbf{e}_{p}(\omega_{p})\times\mathbf{h}_{p}^{*}(\omega_{p})\right]\big\vert
dS\right)^{2}}{\displaystyle \int_{S_{\mathrm{nl}}}\big\vert
\mathfrak{Re}\left[\mathbf{e}_{p}(\omega_{p})\times\mathbf{h}_{p}^{*}(\omega_{p})\right]\big\vert^{2}
dS}.
\end{equation}

The types of nonlinear interactions incorporated in our theoretical model described by
Eqs.~\eqref{FWMpsi} are summarized in Fig.~\ref{fig:endiagr} \textit{via} the energy diagrams
defined by the frequencies of the specific pairs of interacting photons. Thus, as per
Fig.~\ref{fig:endiagr}(a), the terms proportional to the $\gamma_{\mu}^{\prime}$ and
$\gamma_{\mu}^{\prime\prime}$ coefficients describe SPM and degenerate TPA effects, respectively,
whereas Fig.~\ref{fig:endiagr}(b) illustrates XPM and XAM (also called nondegenerate TPA)
interactions whose strength is proportional to $\gamma_{\mu\nu}^{\prime}$ and
$\gamma_{\mu\nu}^{\prime\prime}$, respectively. Finally, there are two distinct types of FWM
processes, represented in Fig.~\ref{fig:endiagr}(c) and Fig.~\ref{fig:endiagr}(d). In the first
case two pump photons combine and generate a pair of photons, one at the signal frequency and the
other one at the idler, a process described by the term proportional to $\gamma_{psi}$. The
reverse process, represented by the $\gamma_{ips}$ and $\gamma_{spi}$ terms, corresponds to the
case in which a signal and an idler photon combine to generate a pair of photons at the pump
frequency.

As Eqs.~\eqref{FWMcouplcoefflinnlav} show, the linear and nonlinear optical coefficients of the
waveguide depend on the index of refraction of silicon, both explicitly and implicitly
\textit{via} the optical modes of the waveguide. In our calculations the implicit modal frequency
dispersion is not taken into account because it cannot be incorporated in the PWE method used to
compute the modes. On the other hand, the explicit material dispersion is accounted for
\textit{via} the following Sellmeier equation describing the frequency dependence of the index of
refraction of silicon \cite{p98book}:
\begin{equation}
n^2(\lambda)=\epsilon+\frac{A}{\lambda^2}+\frac{B\lambda_1^2}{\lambda_1^2-\lambda^2},
\label{index}
\end{equation}
where $\lambda_1=\SI{1.1071}{\micro\meter}$, $\epsilon=11.6858$,
$A=\SI{0.939816}{\square\micro\meter}$, and $B=\num{8.10461e-3}$.
\begin{figure}[t]
\centerline{\includegraphics[width=8cm]{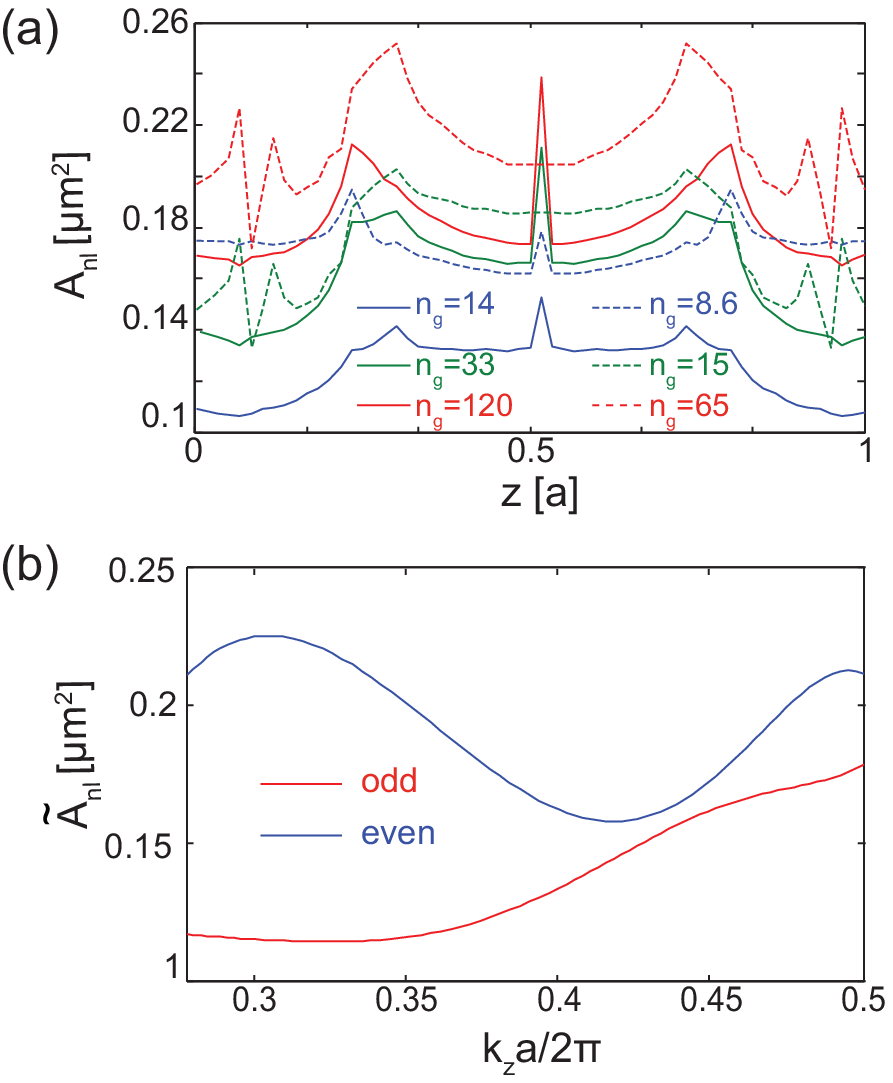}} \caption{(a) Dependence of $A_{\mathrm{nl}}$
on $z$, determined for the odd (solid line) and even (dashed line) modes for several values of the
group-index, $n_{g}$. (b) Frequency dispersion of $\tilde{A}_{\mathrm{nl}}$ calculated for the two
modes, in the spectral domain where they are guiding modes.} \label{fig:area}
\end{figure}

The system of coupled equations, Eqs.~\eqref{FWMpsi} and Eq.~\eqref{FWMFCdyn}, form the basis for
our analysis of degenerate FWM in silicon PhC waveguides. In our simulations, based on numerical
integration of this system of equations using a standard split-step Fourier method combined with a
fifth-order Runge-Kuta method for the integration of the linear, carriers dependent terms, the
$z$-dependence of the coefficients in these equations is rigorously taken into account. However,
one can significantly decrease the simulation time by averaging these fast-varying coefficients
over a lattice constant, as this way the integration step for the resulting, averaged system can
be increased considerably. The derivation of this averaged model is presented in the Appendix.
\begin{figure}[t]
\centerline{\includegraphics[width=8cm]{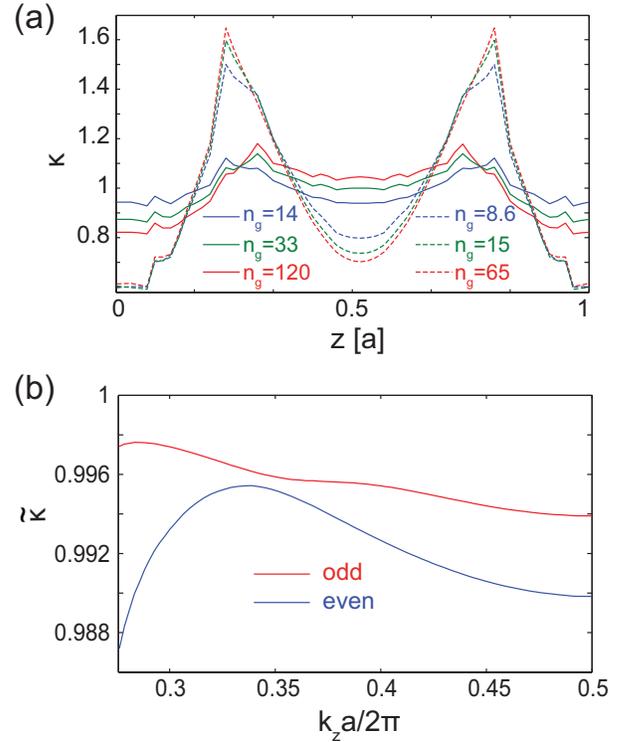}} \caption{(a) Dependence of $\kappa$ on $z$,
determined for the odd (solid line) and even (dashed line) modes for several values of the
group-index, $n_{g}$. (b) Frequency dispersion of $\tilde{\kappa}$ calculated for the two modes,
in the spectral domain where they are guiding modes.} \label{fig:kappa}
\end{figure}

One of the key differences between our theoretical description of FWM processes in Si-PhCWGs and
the widely used models for FWM in waveguides with uniform cross-section, such as optical fibers or
silicon photonic wires, is that the linear and nonlinear waveguide coefficients are periodic
functions of the distance along the waveguide. In what follows, we discuss this feature of the FWM
in more detail, starting with the effective area, $A_{\mathrm{nl}}$, defined by
Eq.~\eqref{effAreaSimplSW}. The dependence of this area on the longitudinal distance, $z$, is
presented in Fig.~\ref{fig:area}(a), where $z$ spans the length of a unit cell. As we have
discussed, a physical characteristic of slow-light modes is their increased spatial extent. This
property is clearly illustrated in Fig.~\ref{fig:area}(a), which shows that in the case of the
even and odd modes the effective area increases by almost a factor of two when the group index
varies from \numrange{14}{120} and from \numrange{8.6}{65}, respectively. This property is also
illustrated by the frequency dispersion of the effective area, averaged over a unit cell, as per
Fig.~\ref{fig:area}(b). Thus, it can be seen in this figure that the effective area has a maximum
at $k_z\approx 0.3(2\pi/a)$ for the even mode and at the edge of the Brillouin zone for both
modes, namely in the regions of slow light indeed.
\begin{figure}[t]
\centerline{\includegraphics[width=8cm]{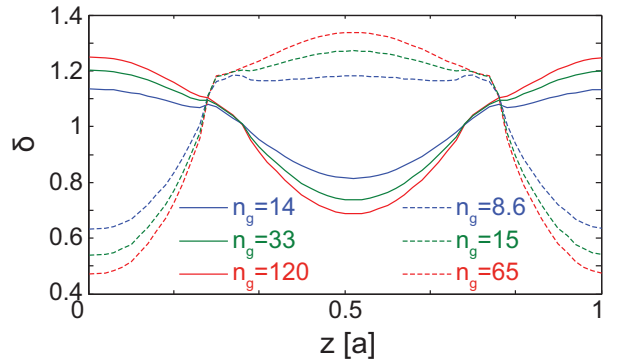}} \caption{Dependence of $\delta$ on $z$,
determined for several values of $n_{g}$. Solid and dashed lines correspond to the odd and even
mode, respectively.} \label{fig:delta}
\end{figure}

The $z$-dependence of the spatial mode overlap, $\kappa$, and the frequency dispersion of its
spatial average over a unit cell, $\tilde{\kappa}$, are plotted in Figs.~\ref{fig:kappa}(a) and
\ref{fig:kappa}(b), respectively. These figures show that the mode overlap varies more strongly
with $z$ in the case of the even mode, whereas in both cases the mode overlap variation increases
as the group-index, $n_{g}$, increases. Interestingly enough, the averaged overlap coefficient of
the even mode has a maximum at $k_z\approx 0.3(2\pi/a)$, i.e.
$\lambda\approx\SI{1.52}{\micro\meter}$, which coincides with a minimum of its $v_{g}$. Note also
that whereas $\kappa(z)$ can be larger than unity within the unit cell, its average,
$\tilde{\kappa}<1$. This result is expected because $\tilde{\kappa}$ quantifies the mode overlap
with the slab waveguide.

\begin{figure}[t]
\centerline{\includegraphics[width=8cm]{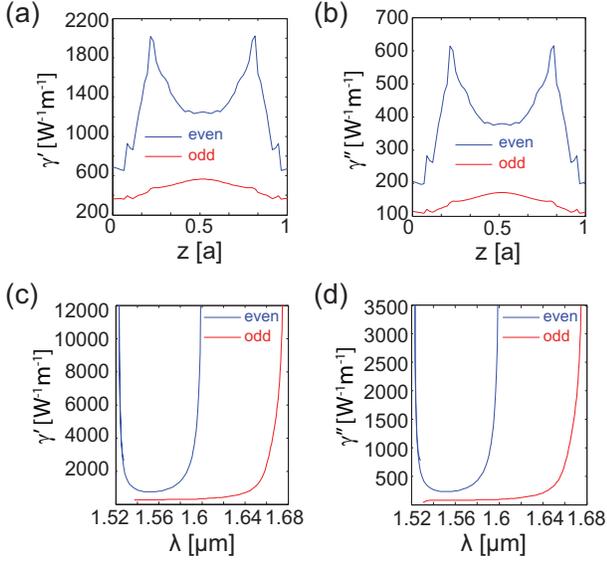}} \caption{(a), (b) Dependence of
$\gamma^{\prime}$ and $\gamma^{\prime\prime}$ on $z$, respectively, determined for
$k_z=0.35(2\pi/a)$. (c), (d) Frequency dispersion of spatially averaged values of
$\gamma^{\prime}(z)$ and $\gamma^{\prime\prime}(z)$, respectively, determined both for the even
and odd modes.} \label{fig:gamma}
\end{figure}
In Fig.~\ref{fig:delta} we present the dependence on $z$ of another physical quantity that
characterizes the linear optical properties of the PhC waveguide, namely its dispersive
properties. This parameter, $\delta$, quantifies the extent to which the $z$-dependent dispersion
coefficients differ from their averaged values. Similarly to the mode overlap coefficient
$\kappa$, $\delta(z)$ shows a more substantial changes with $z$ in the case of the even mode as
compared to the odd one and an increase of the amplitude of these oscillations with the increase
of $n_{g}$. Moreover, as it has been demonstrated in the preceding section, the average of
$\delta(z)$ over a unit cell is equal to unity.

The $z$-dependence of the nonlinear waveguide coefficient that characterizes the strength of SPM
and TPA effects and the wavelength dependence of its average over a unit cell are plotted in the
top and bottom panels of Fig.~\ref{fig:gamma}, respectively. One relevant result illustrated by
these plots is that the nonlinear waveguide coefficient increases considerably as the GV of the
optical mode is tuned to the slow-light regime. Indeed, for the case presented in
Figs.~\ref{fig:gamma}(a) and \ref{fig:gamma}(b) the group-index of the odd and even modes are
$n_{g}=6.4$ and $n_{g}=11.5$, respectively. This phenomenon is better illustrated by the
wavelength dependence of the spatially averaged values of $\gamma^{\prime}(z)$ and
$\gamma^{\prime\prime}(z)$, which are shown in Figs.~\ref{fig:gamma}(c) and \ref{fig:gamma}(d),
respectively. Thus, these plots indicate that the nonlinear waveguide coefficient increases by
more than an order of magnitude as the wavelength is tuned from the fast-light to the slow-light
regime, the nonlinear interactions being enhanced correspondingly.

\section{Phase-matching condition}\label{sPMC}
Before we use the theoretical model we developed to investigate the properties of FWM in
Si-PhCWGs, we derive and discuss the conditions in which optimum nonlinear pulse interaction can
be achieved. In particular, efficient FWM is achieved when the interacting pulses are
phase-matched, namely when the total (linear plus nonlinear) wavevector mismatch is equal to zero.
In the most general case this phase-matching condition depends in an intricate way on the peak
power of the pump, $P_{p}$, signal, $P_{s}$, and idler, $P_{i}$, as well as on the linear and
nonlinear coefficients of the waveguide \cite{a13book}. This complicated relation takes a very
simple form when one considers an experimental set-up most used in practice, namely when the pump
is much stronger than the signal and idler, $P_{p}\gg P_{s},P_{i}$. Under these circumstances, the
phase-matching condition can be expressed as:
\begin{equation}
\label{phase1} 2\gamma_{p}^{\prime} P_p-2\beta_p+\beta_s+\beta_i=0.
\end{equation}

In order to determine the corresponding wavelengths of the optical pulses, this relation must be
used in conjunction with the energy conservation relation, that is $2\omega_p=\omega_s+\omega_i$.

An alternative phase-matching condition, less accurate but easier to use in practice, can be
derived by expanding the propagation constants, $\beta_{s,i}(\omega)$, in Taylor series around the
pump frequency, $\omega_p$:

\begin{figure}[t]
\centerline{\includegraphics[width=8cm]{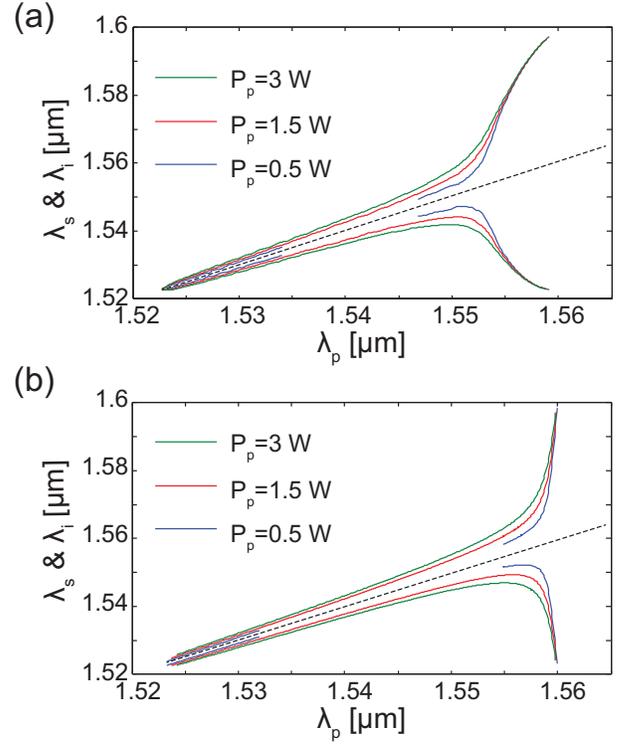}} \caption{(a), (b) Wavelength diagrams
defined by Eq.~\eqref{phase1} and Eq.~\eqref{phase2}, respectively. In both panels dashed lines
correspond to $\omega_p=\omega_{s}=\omega_{i}$.} \label{fig:wavediagr}
\end{figure}
\begin{align}
\label{taylor}
\beta_{s,i}(\omega)&=\sum_{n\geq0}\frac{(\omega-\omega_p)^{n}}{n!}\left.\frac{d^{n}\beta_{s,i}}{d\omega^{n}}\right|_{\omega=\omega_{p}}.
\end{align}
Inserting these expressions in Eq.~\eqref{phase1} and neglecting all terms beyond the
fourth-order, one arrives to the following relation:
\begin{equation}
\label{phase2}
2\gamma_{p}^{\prime}P_p+\beta_{2p}(\Delta\omega)^2+\frac{1}{12}\beta_{4p}(\Delta\omega)^4=0,
\end{equation}
where $\Delta \omega\equiv\vert\omega_p-\omega_s\vert=\vert\omega_p-\omega_i\vert$.

The wavelength diagrams presented in Figs.~\ref{fig:wavediagr}(a) and \ref{fig:wavediagr}(b)
display the triplets of wavelengths for which the phase-matching conditions expressed by
Eq.~\eqref{phase1} and Eq.~\eqref{phase2}, respectively, are satisfied. These wavelength diagrams
were calculated only for the even mode because only this mode possesses spectral regions with
anomalous dispersion [cf. Fig.~\ref{fig:disp}(b)], which is a prerequisite condition for
phase-matching the FWM. More specifically, efficient FWM can be achieved if the pump wavelength
ranges from $\lambda_p=$~\SIrange{1.52}{1.56}{\micro\meter}. Moreover, the diagrams in
Fig.~\ref{fig:wavediagr} show that the predictions based on Eq.~\eqref{phase1} and
Eq.~\eqref{phase2} are in good agreement, especially when $\Delta\omega$ is small. They start to
agree less as $\Delta\omega$ increases because the contribution of the terms discarded when the
series expansion of $\beta_{s,i}(\omega)$ is truncated increases as $\Delta\omega$ increases.

Figure~\ref{fig:wavediagr} also suggests that the spectral domain in which efficient FWM is
achieved depends on the pump power, $P_{p}$. To be more specific, it can be seen that for
$P_{p}\lesssim\SI{0.7}{\watt}$, a spectral gap opens where the phase-matching condition cannot be
satisfied. The spectral width of this gap increases when $P_{p}$ decreases as the the waveguide
was not designed to possess phase-matched modes in the linear regime. Moreover, the diagrams
presented in Fig.~\ref{fig:wavediagr} show that in the fast-light regime the wavelengths defined
by the phase-matching condition depend only slightly on $P_{p}$, whereas a much stronger
dependence is observed when the wavelengths of the signal and idler lie in slow-light spectral
domains.

\section{Results and discussion}\label{sResults}
In this section we illustrate how our theoretical model can be used to investigate various
phenomena related to FWM in Si-PhCWGs. In particular, we will compare the pulse interaction in
slow- and fast-light regimes, calculate the FWM gain, and investigate the influence of various
waveguide parameters on the FWM process. The choice of the values of physical parameters of the
co-propagating pulses and that of the input pump power has been guided by the exact phase-matching
condition given by Eq.~\eqref{phase1}. In all our calculations we assumed that the pulses
propagate in the even mode and, unless otherwise specified, the following values for the pulse and
waveguide parameters have been used in all our simulations: the input peak pump power,
$P_p=10^2P_s=\SI{5}{\watt}$, the input pulse width, $T_p=T_{s}=\SI{7}{\pico\second}$, and the
intrinsic waveguide loss coefficient, $\alpha_{\mathrm{in}}=\SI{50}{\decibel\per\centi\meter}$.

\begin{figure}[t]
\centerline{\includegraphics[width=8cm]{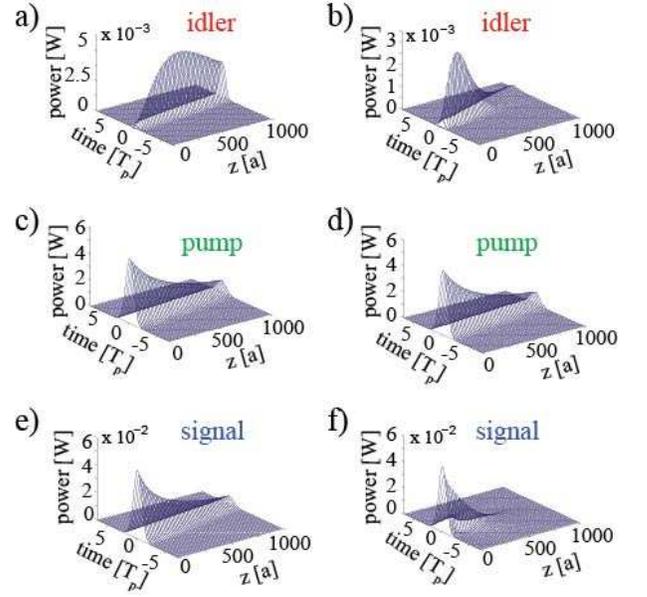}} \caption{Pulse evolution in the time
domain. Left (right) panels correspond to fast-light (slow-light) regimes, the group-index of the
pulses being: $n_{g,i}=9.48$ ($n_{g,i}=20.3$), $n_{g,p}=8.64$ ($n_{g,p}=8.69$), $n_{g,s}=10.37$
($n_{g,s}=23.3$).} \label{fig:pulsetime}
\end{figure}
Let us consider first the evolution of the envelopes of the pulses in the time domain, both in the
slow- and fast-light regimes, as illustrated in Fig.~\ref{fig:pulsetime}. The triplet of
wavelengths for which the phase-matching condition is satisfied is
$\lambda_p=\SI{1554}{\nano\meter}$, $\lambda_s=\SI{1536}{\nano\meter}$, and
$\lambda_i=\SI{1571}{\nano\meter}$ in the fast-light regime, whereas in the slow-light regime the
wavelengths are $\lambda_p=\SI{1559}{\nano\meter}$, $\lambda_s=\SI{1524}{\nano\meter}$, and
$\lambda_i=\SI{1597}{\nano\meter}$. We stress that in both cases the pump pulse propagates in the
fast-light regime, whereas the signal and idler are both generated either in the fast- or
slow-light regime.

Under these circumstances, one expects that the pump evolution in the time domain is similar in
the two cases, a conclusion validated by the plots shown in Figs.~\ref{fig:pulsetime}(c) and
\ref{fig:pulsetime}(d). However, the dynamics of the signal and idler are strikingly different
when they propagate in the slow-light or fast-light regimes. There are several reasons that
account for these differences. First, whereas the FCA coefficient, $\alpha_{\mathrm{fc}}$, has
similar values in the two cases, the FCA and intrinsic losses are much larger in the slow-light
regime because the strength of both these effects is inverse proportional to $v_{g}$. This is
reflected in Fig.~\ref{fig:pulsetime} as a much more rapid decay in the slow-light regime of the
signal and idler pulses. Second, it can be seen that in the slow-light regime the idler pulse
grows at a faster rate. This is again a manifestation of slow-light effects. In particular, the
nonlinear coefficient $\gamma_{ips}$, which determines the FWM gain, is inverse proportional to
$(v_{g,i}v_{g,s})^{1/2}$ [see Eq.~\eqref{FWMipsGammaFWMav}]. As a consequence, the FWM gain is
strongly enhanced when both the signal and idler propagate in the slow-light regime.

\begin{figure}[t]
\centerline{\includegraphics[width=8cm]{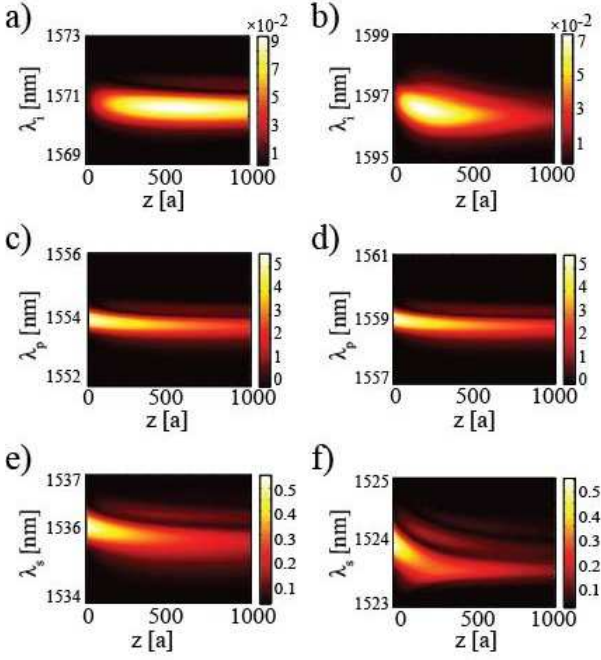}} \caption{From top to bottom, the left
(right) panels show the evolution of the spectra of the idler, pump, and signal in the case of
fast-light (slow-light) regimes. The waveguide and pulse parameters are the same as in
Fig.~\ref{fig:pulsetime}.} \label{fig:pulsespctr}
\end{figure}
The slow-light effects are reflected not only in the characteristics of the time-domain
propagation of the pulses but they also affect the evolution of the pulse spectra. In order to
illustrate this idea, we plot in Fig.~\ref{fig:pulsespctr} the $z$-dependence of the spectra of
the pulses. Similarly to the time-domain dynamics, the spectra of the pump are almost the same in
the slow- and fast-light regimes as its GV does not differ much between the two cases. The most
noteworthy differences between the slow- and fast-light scenarios can again be observed in the
case of the idler and signal. Thus, in the slow-light regime the idler decays faster due to
increased losses and grows and broadens more significantly because of enhanced FWM gain and FCD
effects, respectively. The influence of FCD on the spectral features of the pulses can also be
seen in the case of the signal and, to a smaller extent, the pump. More specifically,
Eq.~\eqref{nFC} shows that the index of refraction of the waveguide decreases due to the
generation of FCs. This in turn leads to a phase-shift and, consequently, a blue-shift of the
pulse \cite{bkr04oe}. Interestingly enough, one can also see in Fig.~\ref{fig:pulsespctr} that as
the frequency of the pulses shifts during their propagation new spectral peaks are forming at the
initial wavelengths for which the phase-matching condition was satisfied.

\begin{figure}[t]
\centerline{\includegraphics[width=8cm]{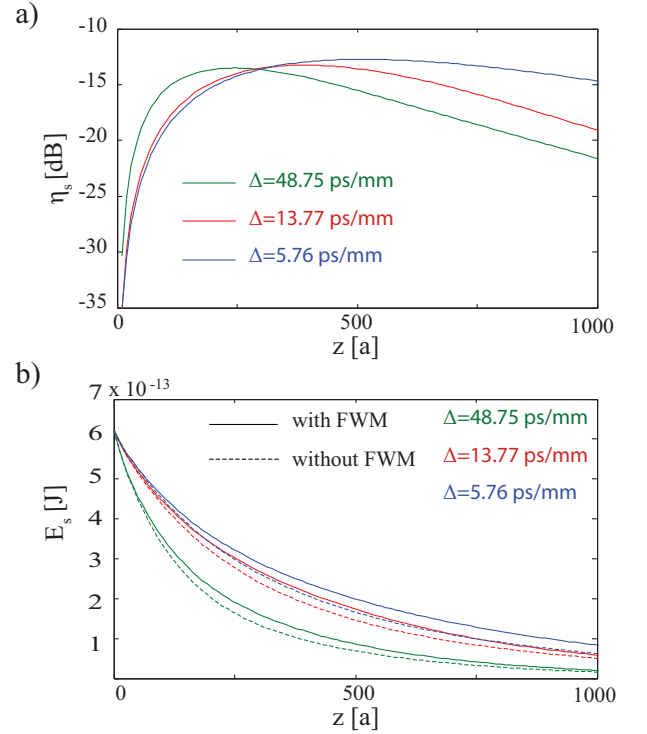}} \caption{(a) FWM enhancement factor vs.
propagation distance, determined for different values of the walk-off parameter,
$\Delta=1/v_{g,s}-1/v_{g,p}$. (b) Signal energy vs. propagation distance, calculated by including
FWM terms in Eqs.~\eqref{FWMpsi} and Eq.~\eqref{FWMFCdyn} and by setting them to zero. The blue
(green) curve corresponds to the fast-light (slow-light) regime considered in
Fig.~\ref{fig:pulsetime}(a) [Fig.~\ref{fig:pulsetime}(b)], whereas the remaining triplet of
phase-matched wavelengths is $\lambda_p=\SI{1556}{\nano\meter}$,
$\lambda_s=\SI{1530}{\nano\meter}$, and $\lambda_i=\SI{1582}{\nano\meter}$ (red).}
\label{fig:gain}
\end{figure}
In order to gain a deeper insight into the influence of slow-light effects on the FWM process, we
computed the $z$-dependence of the pulse energies when the frequencies of the signal and idler
were tuned in the slow-light regions of the even mode of the waveguide. We considered two
scenarios, namely these energies were calculated by including FWM terms in Eqs.~\eqref{FWMpsi} and
Eq.~\eqref{FWMFCdyn} and, in the other case, by setting them to zero, that is
$\gamma_{psi}=\gamma_{spi}=\gamma_{ips}=0$. In the former case, the FWM terms are responsible for
transferring energy from the pump pulse to the signal and idler. Therefore, a suitable quantity to
characterize the efficiency of this energy transfer is what we call the FWM enhancement factor,
$\eta$, which in the case of the signal is defined as
$\eta_{s}=10\log[(E_{SXF}-E_{SX})/E_{s,in}]$. Here, $E_{SXF}$ and $E_{SX}$ are the signal energies
calculated by taking into account, in one case, SPM, XPM, and FWM effects, and only SPM and XPM
terms in the other case (i.e. FWM terms are neglected in the latter case), and $E_{s,in}$ is the
input energy of the signal.

The results of these calculations are summarized in Fig.~\ref{fig:gain}. In particular, it can be
clearly seen in Fig.~\ref{fig:gain}(a) that the FWM enhancement factor is strongly dependent on
pulse propagation regime. To be more specific, as the signal and idler are shifting in the
slow-light regime a smaller amount of energy is transferred from the pump pulse to the signal.
There are two effects whose combined influence leads to this behavior. First, as we discussed, the
pulses experience larger optical losses in the slow-light regime and therefore the signal losses
energy at higher rate. Equally important, as the pulses are tuned in the slow-light regime the
walk-off parameter, $\Delta$, defined as $\Delta=1/v_{g,s}-1/v_{g,p}$, increases, meaning that the
pulses interact for a shorter time and consequently less energy is transferred to the signal.
These conclusions are clearly validated by the results summarized in Fig.~\ref{fig:gain}(b), where
we plot the energy of the signal vs. the propagation distance, determined for several values of
the walk-off parameter. In addition, this figure somewhat surprisingly suggests that the FWM
process is more efficient in the fast-light regime, which is again due to the fact that the pump
and signal overlap over longer time.

\begin{figure}[t]
\centerline{\includegraphics[width=8cm]{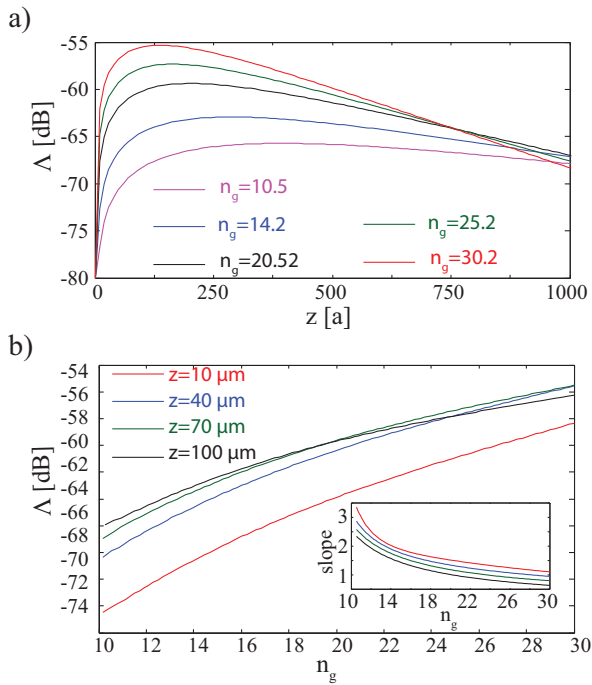}} \caption{(a) Dependence of loss factor,
$\Lambda$, on the propagation distance, $z$, determined for different values of the group-index,
$n_{g}$. (b) Dependence of $\Lambda$ on $n_{g}$, determined for different values of $z$. The
slopes of the curves corresponding to $z=\SI{10}{\micro\meter}$ and $z=\SI{70}{\micro\meter}$ are
shown in the inset.} \label{fig:loss}
\end{figure}
It is well known that in the slow-light regime linear optical effects are enhanced by a factor of
$c/v_{g}$, whereas cubic nonlinear interactions increase by a factor of $(c/v_{g})^{2}$. For
example, FCA and TPA are proportional to $v_{g}^{-1}$ and $v_{g}^{-2}$, respectively. Our
theoretical model predicts, however, that when the mutual interaction between FCs and the optical
field is taken into account these scaling laws can significantly change. This can be understood as
follows: the amount of FCs generated via TPA, $N$, is proportional to $v_{g}^{-2}$ and since FCA
is proportional to the product $v_{g}^{-1}N$, it scales with the GV as $v_{g}^{-3}$.

In order to validate this argument we have determined the optical total loss experienced by a
pulse when it propagates in the presence of TPA and FCA or, in a different scenario, when only TPA
is present [the latter case is realized by simply setting $\alpha_{\mathrm{fc}}=0$ in
Eqs.~\eqref{FWMpsi}]. Moreover, to simplify our analysis, we consider the propagation of only one
pulse by setting all parameters describing XPM and FWM interactions to zero. Finally, we also
reduced the input power to \SI{100}{\milli\watt} in order to avoid strong SPM-induced pulse
reshaping. Under these conditions, the effect of the FCA on the pulse dynamics can be conveniently
characterized by introducing a loss factor, $\Lambda$, defined as
$\Lambda=10\log[(E_{T}-E_{TF})/E_{in}]$, where $E_{TF}$ and $E_{T}$ are the pulse energies in the
case when both TPA and FCA terms are included in the model and when only TPA is present,
respectively, and $E_{in}$ is the input energy of the pulse. The results of these calculations are
presented in Fig.~\ref{fig:loss}.

The variation of the loss factor, $\Lambda$, with the propagation distance, determined for several
values of the GV is presented in Fig.~\ref{fig:loss}(a). As one would have expected, the loss
factor increases with the group-index, $n_{g}$, which is a reflection of the fact that the
FC-induced losses increase with the decrease of the GV. One can observe, however, that when the
propagation distance is larger than about $139a$ the loss factor begins to decrease when the GV
decreases. This behavior is a direct manifestation of slow-light effects, namely as the frequency
is tuned to the slow-light regime the optical losses increase significantly irrespective of the
fact that only TPA is considered or both TPA and FCA effects are incorporated in the numerical
simulations.

This subtle dependence of FC-induced losses on $v_{g}$ is perhaps better reflected by the plots
presented in Fig.~\ref{fig:loss}(b). Thus, for several values of the propagation distance, we have
determined the variation of $\Lambda$ with the group-index, $n_{g}$. Then, by calculating the
slope of the function $\Lambda(n_{g})$ represented on a logarithmic scale one can determine how
FC-induced losses scale with $v_{g}$ [cf. the inset in Fig.~\ref{fig:loss}(b)]. The results of
this analysis clearly demonstrate that FC losses are proportional to $v_{g}^{-3}$, which agrees
with the predictions of our qualitative evaluation of this dependence. We stress that for large
$n_{g}$ (i.e., small $v_{g}$) the $v_{g}^{-3}$ dependence no longer holds at large propagation
distance, chiefly because the pulse is strongly reshaped in the slow-light regime due to enhanced
nonlinear optical effects, and thus its peak power is no longer exclusively determined by optical
losses.

\section{Conclusion}\label{sConcl}
In conclusion, we have derived a rigorous theoretical model, which describes pulsed
four-wave-mixing in one-dimensional photonic crystal slab waveguides made of silicon. Our
theoretical model rigorously incorporate all key linear and nonlinear optical effects affecting
the optical pulse dynamics, including modal dispersion, free-carrier dispersion, free-carrier
absorption, self- and cross-phase modulation, two-photon absorption, cross-absorption modulation,
and four-wave mixing. In addition, the mutual interaction between photogenerated free-carriers and
optical field is incorporated in our theoretical analysis in a natural way by imposing the
conservation the total energy of the optical field and free-carriers. Importantly, our theoretical
formalism allows one to derive rigorous formulae for the optical coefficients characterizing the
linear and nonlinear optical properties of the photonic crystal waveguides, avoiding thus any of
the approximations that are commonly used in the investigation of nonlinear pulse dynamics in
semiconductor waveguides based on photonic crystals.

As a practical application of the theoretical results developed in this study, we have used our
theoretical model to investigate the properties of degenerate four-wave-mixing of optical pulses
propagating in photonic crystal waveguides made of silicon, with a special focus being on
highlighting the differences between the pulse dynamics in the slow- and fast light regimes. This
analysis has revealed not only that linear and nonlinear effects are enhanced in the slow-light
regime by a factor of $n_{g}$ and $n_{g}^{2}$, respectively, but also that these scaling laws are
markedly affected in the presence of free-carriers. Moreover, since our study has been performed
in a very general framework, i.e. generic optical properties of the waveguides (multi-mode
waveguides) and pulse configuration (multi-frequency optical field), our findings can also be used
to describe many phenomena not considered in this work. For example, important nonlinear effects,
including stimulated and spontaneous Raman scattering, coherent anti-Stokes Raman scattering, and
third-harmonic generation, can be included in our model by simply adding the proper nonlinear
polarizations.

\begin{acknowledgments}
This work was supported by the Engineering and Physical Sciences Research Council, grant No
EP/J018473/1. The work of S. L. was supported through a UCL Impact Award graduate studentship. N.
C. P. acknowledges support from European Research Council / ERC Grant Agreement no.
ERC-2014-CoG-648328.
\end{acknowledgments}

\section*{Appendix: Averaged model describing degenerate four-wave-mixing}\label{sAppendix}
The spatial scale over which the envelope of picosecond pulses varies is much larger than the
lattice constant of the PhC and therefore for such optical pulses one can simplify the system of
equations governing the pulse interaction, i.e. Eqs.~\eqref{FWMpsi} and  Eq.~\eqref{FWMFCdyn}, by
taking the average over one lattice constant. Under these conditions, the corresponding system of
coupled equations can be cast in the following form:
\begin{subequations}\label{FWMpsiAv}
\begin{align}\label{FWMpAv}
i&\bigg(\frac{\displaystyle \partial A_{p}}{\displaystyle \partial z}
+\frac{1}{v_{g,p}}\frac{\partial A_{p}}{\partial
t}\bigg)-\frac{\bar{\beta}_{2,p}}{2}\frac{\partial^{2} A_{p}}{\partial
t^{2}}+\frac{\omega_{p}\delta n_{\mathrm{fc}}\tilde{\kappa}_{p}}{n v_{g,p}}A_{p}\notag \\
&+\frac{ic\tilde{\kappa}_{p}}{2n v_{g,p}}(\alpha_{\mathrm{fc}}+\alpha_{\mathrm{in}})A_{p}+2e^{i\Delta\bar{\beta}z}\tilde{\gamma}_{psi}A_{s}A_{i}A_{p}^{*} \notag \\
&+\left(\tilde{\gamma}_{p}\vert A_{p}{\vert}^{2}+2\tilde{\gamma}_{ps}\vert
A_{s}{\vert}^{2}+2\tilde{\gamma}_{pi}\vert A_{i}{\vert}^{2}\right)A_{p}=0, \\
\label{FWMsAv}i&\bigg(\frac{\displaystyle
\partial A_{s}}{\displaystyle \partial z} +\frac{1}{v_{g,s}}\frac{\partial
A_{s}}{\partial t}\bigg)-\frac{\bar{\beta}_{2,s}}{2}\frac{\partial^{2}
A_{s}}{\partial t^{2}}+\frac{\omega_{s}\delta n_{\mathrm{fc}}\tilde{\kappa}_{s}}{n v_{g,s}}A_{s}\notag \\
&+\frac{ic\tilde{\kappa}_{s}}{2n v_{g,s}}(\alpha_{\mathrm{fc}}+\alpha_{\mathrm{in}})A_{s}+e^{-i\Delta\bar{\beta}z}\tilde{\gamma}_{spi}A_{p}^{2}A_{i}^{*} \notag \\
&+\left(\tilde{\gamma}_{s}\vert A_{s}{\vert}^{2}+2\tilde{\gamma}_{sp}\vert
A_{p}{\vert}^{2}+2\tilde{\gamma}_{si}\vert A_{i}{\vert}^{2}\right)A_{s}=0, \\
\label{FWMiAv}i&\bigg(\frac{\displaystyle
\partial A_{i}}{\displaystyle \partial z} +\frac{1}{v_{g,i}}\frac{\partial
A_{i}}{\partial t}\bigg)-\frac{\bar{\beta}_{2,i}}{2}\frac{\partial^{2}
A_{i}}{\partial t^{2}}+\frac{\omega_{i}\delta n_{\mathrm{fc}}\tilde{\kappa}_{i}}{n v_{g,i}}A_{i}\notag \\
&+\frac{ic\tilde{\kappa}_{i}}{2n v_{g,i}}(\alpha_{\mathrm{fc}}+\alpha_{\mathrm{in}})A_{i}+e^{-i\Delta\bar{\beta}z}\tilde{\gamma}_{ips}A_{p}^{2}A_{s}^{*} \notag \\
&+\left(\tilde{\gamma}_{i}\vert A_{i}{\vert}^{2}+2\tilde{\gamma}_{ip}\vert
A_{p}{\vert}^{2}+2\tilde{\gamma}_{is}\vert A_{s}{\vert}^{2}\right)A_{i}=0,
\end{align}
\end{subequations}
\begin{align}\label{FWMFCdynAv}
&\frac{\partial N}{\partial t} = -\frac{N}{\tau_{c}}+\frac{1}{\hbar}
\bigg\{\sum_{\mu=p,s,i}\bigg[\frac{\Upsilon_{\mu}^{\prime\prime}}{\omega_{\mu}}\vert
A_{\mu}{\vert}^{4} \notag \\
&~+\sum\limits_{\scriptsize{\begin{array}{c}
        \nu=p,s,i \\
        \nu\neq \mu
      \end{array}}}\frac{4\Upsilon_{\mu\nu}^{\prime\prime}}{\omega_{\mu}+\omega_{\nu}}
\vert A_{\mu}{\vert}^{2}\vert
A_{\nu}{\vert}^{2}\bigg]+\frac{1}{\omega_{p}}\mathfrak{Im}\big[2\Upsilon_{psi} \notag \\
&~\times {A_{p}^{*}}^{2}A_{s}A_{i}e^{i\Delta\bar{\beta}z}+(\Upsilon_{spi}+\Upsilon_{ips})
A_{p}^{2}A_{s}^{*}A_{i}^{*}e^{-i\Delta\bar{\beta}z}\big]\big\}.
\end{align}
The coefficients of the linear and nonlinear terms in these equations are given by the following
formulae:
\begin{subequations}\label{FWMcouplcoefflinnlAv}
\begin{align}
\label{FWMcouplcoeffLinSAv}
&\tilde{\kappa}_{\mu} = \frac{\epsilon_{0}n^{2}}{2\bar{W}_{\mu}}\int_{V_{\mathrm{nl}}}\vert\mathbf{e}_{\mu}(\omega_{\mu})\vert^{2}dV, \\
\label{FWMGammaSPMAv} &\tilde{\gamma}_{\mu} =
\frac{3\omega_{\mu}\epsilon_{0}a}{16v_{g,\mu}^{2}}\frac{1}{\bar{W}_{\mu}^{2}}\int_{V_{\mathrm{nl}}}\mathbf{e}_{\mu}^{*}(\omega_{\mu}) \notag \\
&~~\cdot\hat{\chi}^{(3)}(\omega_{\mu},-\omega_{\mu},\omega_{\mu})\vdots\mathbf{e}_{\mu}(\omega_{\mu})\mathbf{e}_{\mu}^{*}(\omega_{\mu})\mathbf{e}_{\mu}(\omega_{\mu})dV, \\
\label{FWMGammaXPMdiffFreqAv} &\tilde{\gamma}_{\mu\nu} =
\frac{3\omega_{\mu}\epsilon_{0}a}{16v_{g,\mu}v_{g,\nu}}\frac{1}{\bar{W}_{\mu}\bar{W}_{\nu}}\int_{V_{\mathrm{nl}}}\mathbf{e}_{\mu}^{*}(\omega_{\mu}) \notag \\
&~~\cdot\hat{\chi}^{(3)}(\omega_{\nu},-\omega_{\nu},\omega_{\mu})\vdots\mathbf{e}_{\nu}(\omega_{\nu})\mathbf{e}_{\nu}^{*}(\omega_{\nu})\mathbf{e}_{\mu}(\omega_{\mu})dV, \\
\label{FWMpsiGammaFWMAv} &\tilde{\gamma}_{psi} =
\frac{3\omega_{p}\epsilon_{0}a}{16v_{g,p}{(v_{g,s}v_{g,i})}^{\frac{1}{2}}}\frac{1}{\bar{W}_{p}{(\bar{W}_{s}\bar{W}_{i})}^{\frac{1}{2}}} \notag \\
&~~\times\int_{V_{\mathrm{nl}}}\mathbf{e}_{p}^{*}(\omega_{p})\cdot
\hat{\chi}^{(3)}(\omega_{s},-\omega_{p},\omega_{i})\vdots\mathbf{e}_{s}(\omega_{s})\mathbf{e}_{p}^{*}(\omega_{p})\mathbf{e}_{i}(\omega_{i})dV, \\
\label{FWMspiGammaFWMAv} &\tilde{\gamma}_{spi} =
\frac{3\omega_{s}\epsilon_{0}a}{16v_{g,p}{(v_{g,s}v_{g,i})}^{\frac{1}{2}}}\frac{1}{\bar{W}_{p}{(\bar{W}_{s}\bar{W}_{i})}^{\frac{1}{2}}} \notag \\
&~~\times\int_{V_{\mathrm{nl}}}\mathbf{e}_{s}^{*}(\omega_{s})\cdot
\hat{\chi}^{(3)}(\omega_{p},-\omega_{i},\omega_{p})\vdots\mathbf{e}_{p}(\omega_{p})\mathbf{e}_{i}^{*}(\omega_{i})\mathbf{e}_{p}(\omega_{p})dV, \\
\label{FWMipsGammaFWMAv} &\tilde{\gamma}_{ips} =
\frac{3\omega_{i}\epsilon_{0}a}{16v_{g,p}{(v_{g,s}v_{g,i})}^{\frac{1}{2}}}\frac{1}{\bar{W}_{p}{(\bar{W}_{s}\bar{W}_{i})}^{\frac{1}{2}}} \notag \\
&~~\times\int_{V_{\mathrm{nl}}}\mathbf{e}_{i}^{*}(\omega_{i})\cdot
\hat{\chi}^{(3)}(\omega_{p},-\omega_{s},\omega_{p})\vdots\mathbf{e}_{p}(\omega_{p})\mathbf{e}_{s}^{*}(\omega_{s})\mathbf{e}_{p}(\omega_{p})dV, \\
\label{FWMUpsilonFWMAv}&\Upsilon_{\varpi}=\frac{1}{a}\int_{z_{0}}^{z_{0}+a}\frac{\gamma_{\varpi}(z)}{A_{\mathrm{nl}}(z)}dz.
\end{align}
\end{subequations}
where $V_{\mathrm{nl}}$ is the volume occupied by silicon in a unit cell of the PhC waveguide and
$z_{0}$ an arbitrary distance. Finally, $A_{\mathrm{nl}}(z)$ in Eq.~\eqref{FWMUpsilonFWMAv} is
given by Eq.~\eqref{effAreaSimplSW}, whereas the index $\varpi$ takes any of the following values:
$p$, $s$, $i$, $ps$, $si$, $ip$, $psi$, $spi$, or $ips$.whereas the index $\varpi$ takes any of
the following values: $p$, $s$, $i$, $ps$, $si$, $ip$, $psi$, $spi$, and $ips$. Note that in
deriving the averaged equations governing the pulse and FC dynamics, Eqs.~\eqref{FWMpsiAv} and
Eq.~\eqref{FWMFCdynAv}, we have assumed that the FWM process is nearly phase-matched, namely
$\Delta\bar{\beta}\ll 1/a$. In other words, the phase of the exponential factors in these
equations vary much slower with the distance, $z$, as compared to the variation of the dielectric
constant of the PhC waveguide.

\begin{figure}[t]
\centerline{\includegraphics[width=8cm]{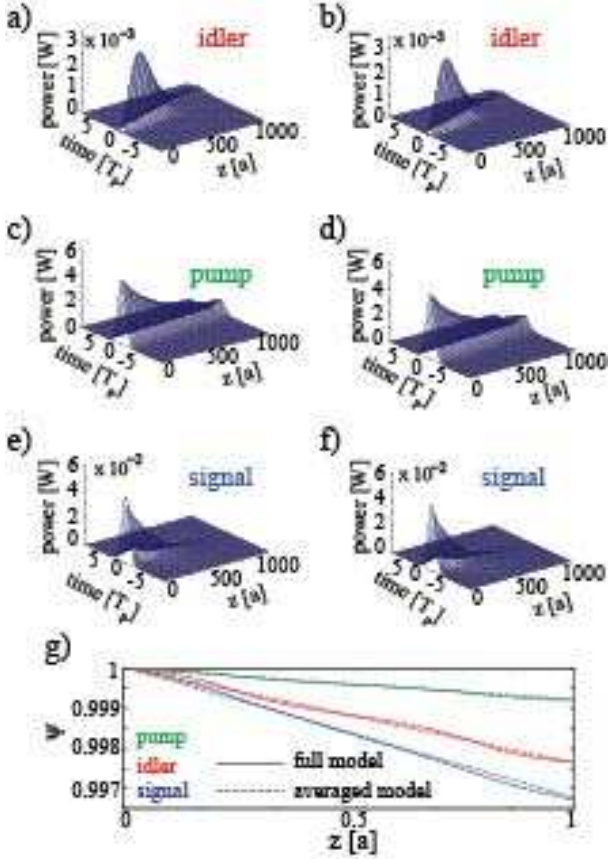}} \caption{Comparison between pulse evolution
as described by the full and averaged model is presented in the left and right panels,
respectively. The group-index of the pulses are $n_{g,i}=20.3$, $n_{g,p}=8.69$, and $n_{g,s}=23.3$
and correspond to the slow-light propagation scenario presented in Fig.~\ref{fig:pulsetime}. The
bottom panel shows the $z$-dependence of the normalized pulse amplitude,
$\Psi_{\mu}(z)=A_{\mu}(z_{0}+z)/A_{\mu}(z_{0})$, $\mu=p,s,i$, calculated for the unit cell
starting at $z_{0}=200a$.} \label{fig:pulsetimecomp}
\end{figure}
A comparison between the predictions of the full and averaged models is illustrated in
Fig.~\ref{fig:pulsetimecomp}. Thus, we have considered the slow-light pulse dynamics presented in
Fig.~\ref{fig:pulsetime} and determined the pulse evolution using both the full and averaged
models. As it can be seen, both models predict a similar pulse dynamics for the entire propagation
length, $z=1000a$. This result is expected as the envelope of picosecond pulses, as are those
chosen in our simulations, spans a large number of unit cells and therefore the pulse amplitude is
only slightly affected by the local inhomogeneity of the index of refraction.

The fast variation with $z$ of the pulse envelope is shown in Fig.~\ref{fig:pulsetimecomp}(g),
where we plot the $z$-dependence of the normalized pulse amplitude,
$\Psi_{\mu}(z)=A_{\mu}(z_{0}+z)/A_{\mu}(z_{0})$, $\mu=p,s,i$, calculated for the unit cell
starting at $z_{0}=200a$. It can be seen in this figure that the pulse envelope varies at a
spatial scale commensurable with the lattice constant yet the amplitude of these variations is
much smaller than the pulse peak amplitude. The magnitude of these variations, however, would
comparatively become more significant should the pulse duration would be brought to the
femtosecond range.

\end{document}